\pdfoutput=1

\documentclass[twocolumn,amssymb,floatfix,deluxe,prd,showkeys]{revtex4}

\usepackage{graphicx}
\usepackage{dcolumn}
\usepackage{bm}
\usepackage{epsfig}

\begin{document}

\title{Two old ways to measure the electron-neutrino mass}

%

\author{A. De R\'ujula}
\affiliation{IFT(UAM), Madrid, Spain; CERN,
1211 Geneva 23, Switzerland}%

\date{\today}

\begin{abstract}
Three decades ago, the measurement of the electron neutrino 
mass in atomic electron capture (EC) experiments was scrutinized
in its two variants: single EC and neutrino-less double EC. 
For certain  isotopes  an
atomic resonance enormously enhances the expected decay rates.
The favoured technique, based on
calorimeters as opposed to spectrometers, has the advantage of greatly
simplifying the theoretical analysis of the data. After an initial surge of measurements,
the EC approach did not seem to be competitive.
But very recently, there has been great progress on micro-calorimeters
and the measurement of atomic mass
differences. Meanwhile, the beta-decay neutrino-mass limits have improved
by a factor of 15, and the difficulty of the experiments by the cube of that figure.
Can the ``calorimetric"  EC theory 
cope with this increased challenge? I answer this question affirmatively.
In so doing I briefly review the subject and extensively address some persistent
misunderstandings of the underlying quantum physics.

\end{abstract}

\maketitle






\section{Motivation}
\label{sec:Motivation}

In 1933 Perrin qualitatively described \cite{Perrin} and Fermi computed
\cite{Fermi} how a nonzero neutrino mass would affect the endpoint of the electron
energies in a $\beta$-decay process. Two score years later, the laboratory quest
for a non-zero result in 
this kind of measurement continues in ernest \cite{Weinheimer}.

In a 1980 paper with almost the same title as this one \cite{ADR} I discussed the possibility 
to measure or constrain the mass of the electron neutrino in various processes involving electron capture (EC). This is
the $e\,p\!\to\! \nu\,n$ weak-interaction process whereby an atomic electron interacts with a 
nucleus of charge $Z$ to produce a neutrino, leaving behind a nucleus of charge $Z-1$
and a hole in the orbital of the daughter atom from which the electron was captured.
There we noted that --if nature was kind enough in choosing the relevant
parameters of some isotopes-- the endpoint counting rates  would be significantly
enhanced by atomic resonances and that ``calorimetric" measurements would obviate
the complications induced by ``atomic and molecular problems".
EC would complement the classical approach  \cite{Perrin,Fermi} employing nuclear 
$\beta$-decay --notably of $^3\rm H$ (Tritium)  and $^{187}\rm Re$--
to constrain the mass of the electron ``anti"-neutrino \cite{Weinheimer}.

Several experiments were performed in the 1980's to study the feasibility of the proposed
method 
with the conclusion that EC was not competitive 
with $^3\rm H$ 
$\beta$-decay, under the eminently reasonable assumption that the two mentioned distinct 
neutrinos have the same mass. Very recently, the hopes concerning electron-capture
experiments have been rekindled,
an opportunity to
manifest glee for the experiments and to review and further complete the theory,
which for a case of current interest: calorimetric measurement in the EC decay of
$^{163}\rm Ho$, was first developed in detail in \cite{ADRML}.

Calorimetry is also being pursued in the $\beta$-decay of
$^{187}\rm Re$ \cite{Rhenium}. Here the theory, as we shall mention, is not as simple
 as for EC decays. One reason why substances as exotic $^3\rm H$, 
$^{187}\rm Re$ and $^{163}\rm Ho$ are employed is that the
total energy released in their decays --the $Q$-values-- are the next to smallest 
or smallest of the periodic table, so that the fraction of
$m_\nu$-sensitive events is largest. Another reason is that their lifetimes
are not prohibitively long.

Neutrino-less double $\beta$ decay is the classic method to attempt to 
establish the Majorana or Dirac nature of massive neutrinos, while
measuring a function of their masses. This process also has
an EC analog: neutrino-less double electron capture \cite{Otros, GGN,BDJ}. There
is also very significant progress concerning the prospects of calorimetry in this field,
mainly involving the decay of $^{152}\rm Gd$ \cite{Eliseev}.

Naturally, the ultimate goal of all the mentioned experiments, and others, is to
compete with the current best laboratory limits on the electron antineutrino mass:
\begin{eqnarray}
m_{\bar \nu} &<& 2.3 \, \rm eV \, (95\% \; C.L.), 
\\
m_{\bar \nu)} &<& 2.05 \, \rm eV \, (95\% \; C.L.) ,
\end{eqnarray}
from the Mainz 
and Troitsk  data,
respectively \cite{Mainz,Troitsk}.

At the time the theories of resonant single \cite{ADRML} and double \cite{BDJ} EC were elaborated,
the limits on --or alleged measurements of-- $m_{\bar \nu}$ were at the 30 eV level. The relative yield of
events sensitive to $m_{\bar \nu}$ or $m_{\nu}$ in a specific decay scales like $m^3$, so
that the experiments have in this sense become approximately $(30/2)^3\sim 3.4\times 10^3$
times more demanding. The question arises whether or not the underlying EC theory is precise
enough to deal with the current experimental situation. The main aim of this paper is to
answer this question --affirmatively-- in the case of the EC and DEC current and planned campaigns.

\section{A trivial reminder}
\label{reminder}

Consider  the $\beta$ decay, $^3\rm{\hat H}\!\to\! ^3\rm{\hat He}\, e^-\, \bar\nu$,
of a free Triton ($^3\rm{\hat H}\!=\! {^3\rm{H}}^+$, $^3\rm{\hat He}\!=\! {^3\rm{He}}^{++}$)
and ignore radiative
corrections, neutrino mixing
and the (negligible) nuclear-recoil effect. Define 
$Q=M_i\!-\!M_f$, with $i,f$ the initial and final nuclei.
Let $F(E_e)$ be the ``Fermi" function reflecting the fact that the electron is born
in a Coulombic field. Let ${\cal M}$ be the $n\!\to\! p\, e\, \bar\nu$ matrix element and $\Phi$
the phase space factor.
Very explicitly, the differential decay width is:
\begin{eqnarray}
{d\Gamma\over dE_e}&=&{1\over 2}\,|{\cal M}|^2\,F(E_e)\,\Phi,
\label{1}\\
|{\cal M}|^2&\approx& 32 \,G_F^2 \cos^2\theta_C\,M_i\,M_f\,(1+3 \,g_A^2\,)E_e\,E_{\bar\nu},
\label{ME}\\
\Phi&=&{1\over 8\,\pi^3}{p_e\,p_{\bar \nu}\over M_i\,M_f},
\label{PS}\\
 p_{\bar \nu}&=&\sqrt{E_{\bar\nu}^2-m_{\bar\nu}^2} \, ,\;\;\;\; E_{\bar\nu}= Q-E_e ,
\label{pE}
\end{eqnarray}
where $g_A\!\approx 1.23$ is the nucleon axial coupling.
In Eq.(\ref{ME}), $|{\cal M}|^2$ takes into account that 
$\{^3\rm{\hat H},{^3}\rm{\hat He}\}$  is an isospin doublet
(misaligned" by $\approx\!\cos\theta_C$ with a weak isodoublet)
so that the Fermi matrix element 
$\langle ^3\rm{\hat He}|\sum_1^3 \tau^+_i|^3\rm{\hat H}\rangle$ is unity.
Similarly, given the simple  structure of these nuclei,
the Gamow-Teller matrix element is
$|\langle ^3\rm{\hat He}|\sum_1^3 \tau^+_i \,\vec\sigma_i |
^{3}\rm{\hat H}\rangle|\simeq \sqrt{3}$,
exact for a free neutron \cite{BW}.

We know from the observations of neutrino oscillations that the electron
neutrino is, to a good approximation, a superposition of three mass 
eigenstates, $\nu_i$: $\nu_e=\sum_i U_{ei} \nu_i$, 
with $\sum_i |U_{ei}|^2= 1$. Thus, we ought to have written $d\Gamma/dE_e$
in Eq.~(\ref{1}) as an incoherent superposition of spectra with weights $|U_{ei}|^2$
and masses $m(\nu_i)$. But the measured differences 
$m^2(\nu_i)-m^2(\nu_j)$ are so small that current direct attempts to measure
the quantity $m_{\bar\nu}$ of Eq.~(\ref{pE}) are certain to reach the required accuracy
only if neutrinos are nearly degenerate in mass, in which case $m_\nu$ in
Eqs.~(\ref{1}-\ref{pE}) stands for their nearly common mass.

The function on which the ``Kurie plots" are based is:
\begin{eqnarray}
K(E_e)&\equiv&
\sqrt{d\Gamma\over F(E_e)\,p_e\,E_e\, dE_e}\propto \sqrt{E_{\bar\nu}\,p_{\bar\nu}}
\nonumber\\
&=&\sqrt{ (Q-E_e)\sqrt{(Q-E_e)^2-m_{\bar\nu}^2}}\, ,
\nonumber
\end{eqnarray} 
a straight line ending at $E_e=Q$ if $m_{\bar\nu}=0$. The spectrum near the endpoint
for $m_{\bar\nu}=0$ is quadratic in $Q-E_e$. The fraction of events potentially
sensitive to $m_{\bar\nu}\neq 0$ --in an interval of width of ${\cal{O}}(m_{\bar\nu})$--
scales as $m_{\bar\nu}^3$, and so do the challenges to experiment and theory.

The neutrino-mass sensitive factor $p_{\bar\nu}$ 
in $K(E_e)$ arises exclusively from phase space,
it depends on the mass difference $Q$ but  is otherwise
independent of the constituency of
the nuclei considered. Going one step inwards in
``resolution", the description of the $p\, e\, \bar\nu$ decay of a free neutron is independent
of the nucleons' constituent quarks and gluons, but for the fact that they determine
(in principle) $m_n$, $m_p$ and  $g_A$.

\begin{figure}
\begin{center}
\includegraphics[width=0.50\textwidth]{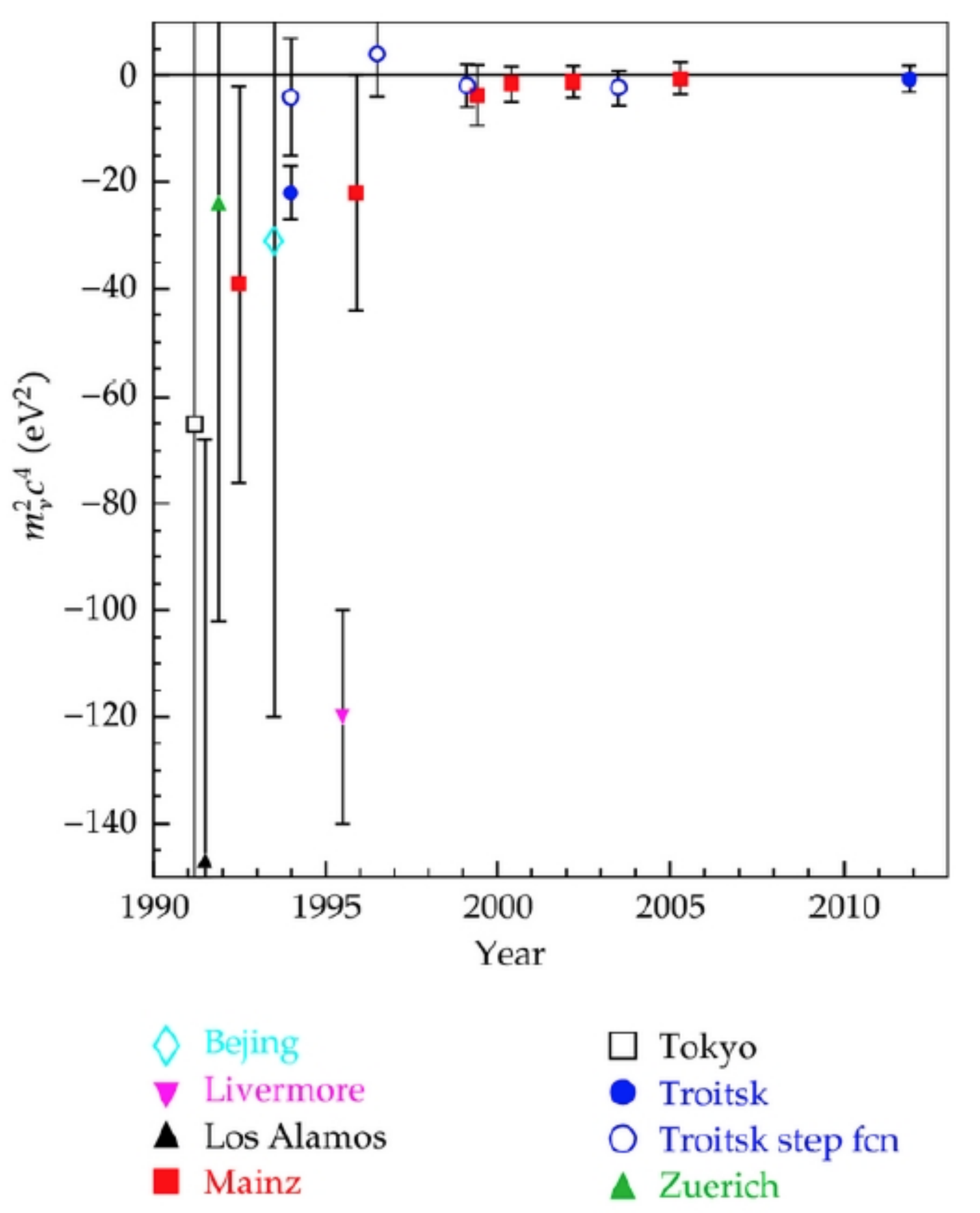}
\vspace{-.5cm}
\caption{Relatively recent progress and some errors in constraining $m_{\bar\nu}$ from 
${^3}\rm H$ $\beta$-decay, as summarized in \cite{Weinheimer}.
\vspace{-1cm}
\label{fig:NuMassHistory}}
\end{center}
\end{figure}

\begin{figure}
\vspace{.3cm}
\hspace{1.8cm}
\begin{center}
\includegraphics[width=0.47\textwidth]{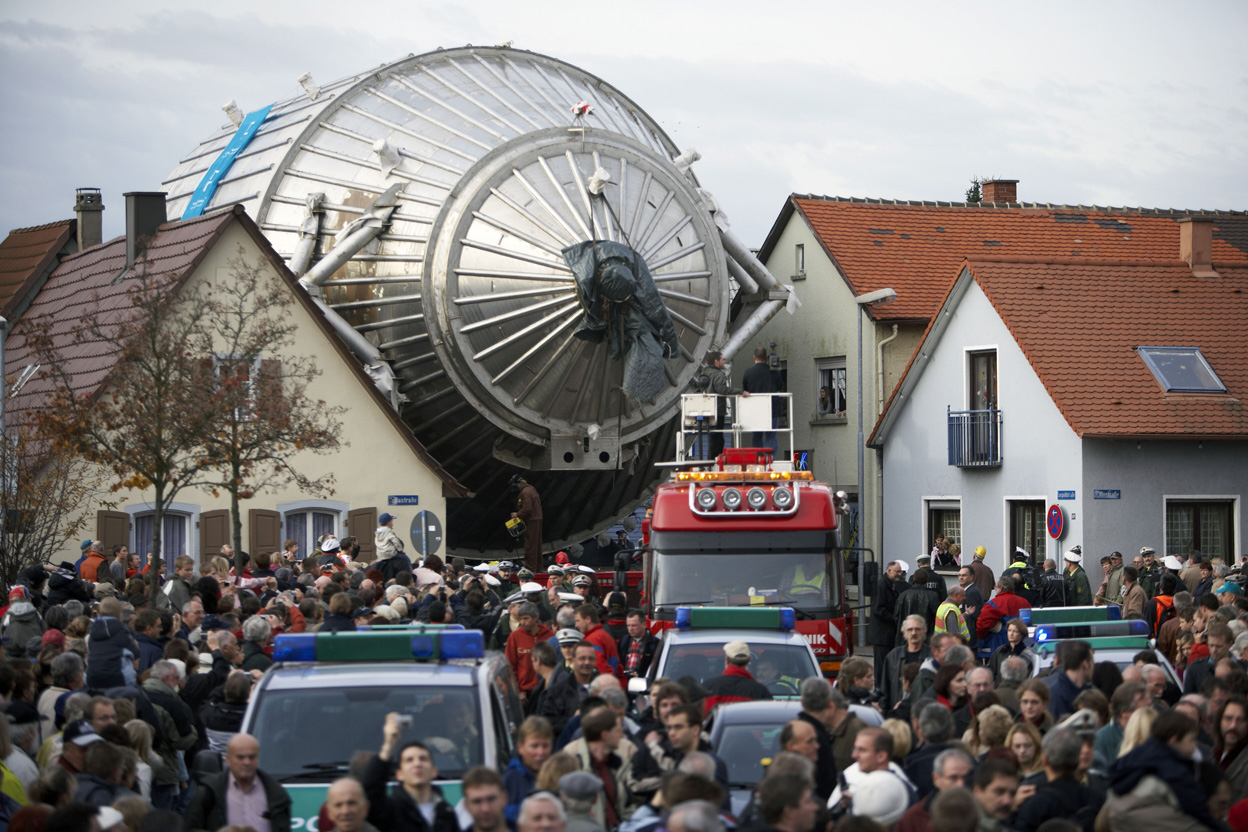}
\caption{The spectrometer of KATRIN in dire straits.
\label{fig:Katrin}}
\end{center}
\end{figure}

The laboratory constraints on $m_{\bar\nu}$ from ${^3}\rm H$ decay have continued
to improve quite impressively in the past two decades, as summarized in
Fig.~(\ref{fig:NuMassHistory}), from \cite{Weinheimer}. The current endeavours are
not trifling, as illustrated in Fig.~(\ref{fig:Katrin}), depicting a toilsome moment
in the transport of  KATRIN's spectrometer to Karlsruhe,
after a 9000 km journey though the Danube, the Black Sea, the Mediterranean,
the Atlantic, the North Sea and the Rhine \cite{Weinheimer}. Across the Atlantic
there is an ongoing test \cite{Formaggio} of  the novel idea behind Project 8: 
to measure single electron energies via their coherent synchrotron radiation \cite{Monreal}.

\section{The Calorimetric ``Principle"}
\label{sec:CP}

\begin{figure}
\hspace{-1.5cm}
\begin{center}
\includegraphics[width=0.47\textwidth]{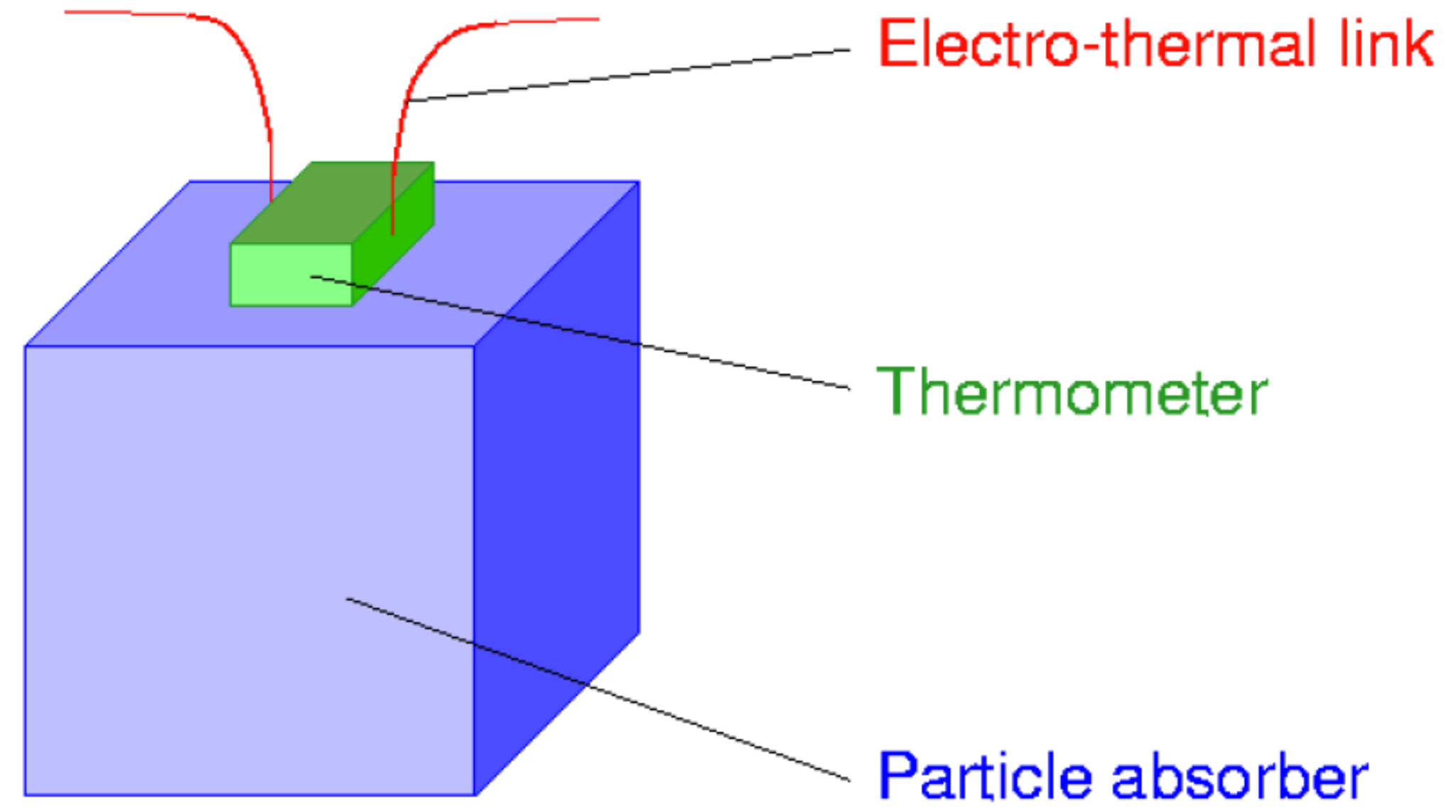}
\caption{A theorists' calorimeter \cite{Weinheimer}. The source is implanted in the absorber,
where all energy but that of escaping neutrinos is deposited and measured as a rise of temperature.
\label{fig:THCalorimeter}}
\end{center}
\end{figure}

A calorimeter, in our discussion, is a detector in which the decaying source is embedded, 
capable of measuring all the energy released in a weak decay, but that of the escaping neutrino.
A view of a calorimeter, presumably meant for theorists, is shown in Fig.~\ref{fig:THCalorimeter}.

Somewhat reminiscent of the M\"ossbauer effect,
the calorimetric ``principle" follows from the realization that one can extend the 
phase-space considerations of \S \ref{reminder}
outwards in resolution, from quarks, nucleons and nuclei all the way to the decay of 
the detector before to the detector after the neutrino (and nothing else) escaped from it. Calling $D_{b,a}$ the detector before and after, and $E_c$
the measured calorimetric energy, the overall process, $D_b\to D_a+E_c+E_\nu$, is
a three-body decay with kinematics --and consequent neutrino-mass sensitivity-- 
as simple as the ones of neutron decay, with $Q$ being now defined as 
$m(D_a)-m(D_b)$.

A problem that has traditionally pestered $^3\rm H$ experiments (almost all of which
were non-calorimetric) is that 
the final atom or molecule may be left in excited states of energy $E_n$ above the ground 
state. 
In a $\beta$-decay experiment in which the electron energy is measured, energy
conservation implies $Q=E_e+E_{\bar \nu}+E_n$, so that the electron spectrum is
a superposition of contributions whose end-points are at $E_e=Q-E_n-m_{\bar\nu}$.
The spectral shape from which $m_{\bar\nu}$ is to be inferred is  a complicated
superposition of spectra with different endpoints.
This is the well known ``atomic or molecular" problem.

The main advantage of a calorimeter is that its measurements are independent of the
various states in which a daughter atom, molecule or crystal may be left, as well as
the various decay channels ($X$-rays or electron-emitting transitions) via which the excited 
final states return to the ground state.
If the de-excitation times, as expected, are much shorter than the  signal's rise-time
all de-excitation energies of a decay event add up in $E_c$. 

The main disadvantage of a calorimeter is that --unlike in a $\beta$-decay spectrometer--
there is no way to veto events whose energy is well below the interesting 
end-point region. The full $E_c$ spectrum is measured. To avoid pile-up (simultaneously
measured events), the activity of the source/detector must be limited. This means that
``calorimeter farms" with up to thousands of micro-calorimeters must be contemplated.
But their individual elements are minute: barely visible to the naked eye!

Strictly speaking, the calorimeter considerations we have
discussed apply only to allowed weak
decays, of which the ${^3}{\rm H}$ and ${^{163}}{\rm Ho}$ cases are examples.
For them the nontrivial nuclear-dependence of the matrix element 
$|{\cal M}|^2$ is just a number, $1+3\,g_A^2$ in $n$ or ${^3}{\rm H}$ decay. In a
forbidden decay such as that of $^{187}\rm Re$, the outgoing $e^-$ or $\bar \nu$
must carry away angular momentum, which induces an $E_e$-dependence of
the corresponding $|{\cal M}|^2$. Though not to leading order, this reintroduces
the need to deal with the different atomic or molecular excitations \cite{Dvornicky}.

\section{ $ \rm{ \bold{{^{187}}Re}}$ and $ \rm{ \bold{{^{163}}Ho}}$ experiments}

\begin{figure}
\hspace{-1.5cm}
\begin{center}
\includegraphics[width=0.50\textwidth]{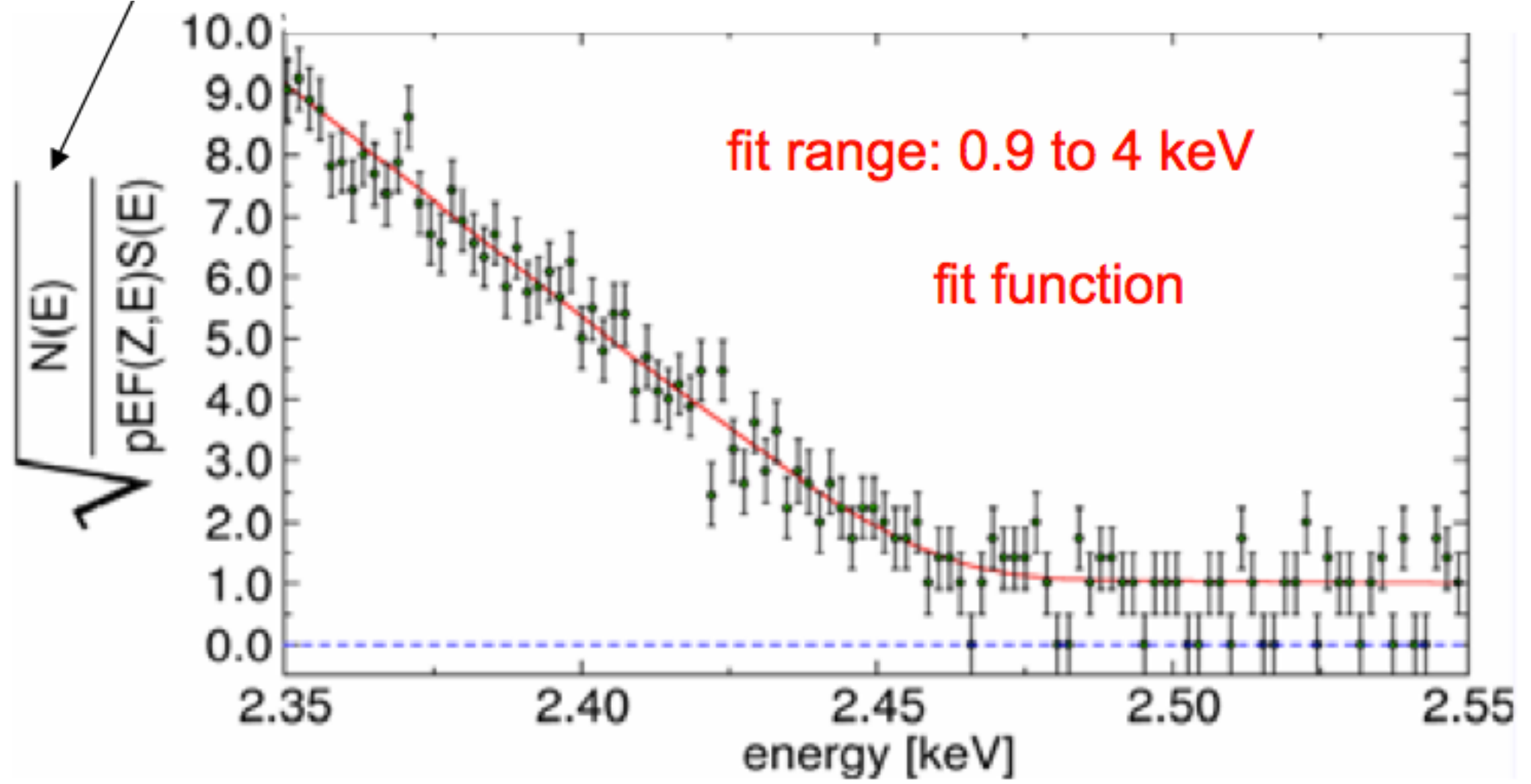}
\caption{A $^{187}\rm Re$ Kurie plot from MIBETA \cite{Rhenium}.
\label{fig:MiBeta}}
\end{center}
\end{figure}

The ground-state to ground-state nuclear
transition $^{187}\rm Re ({5/2})^+$$\,\to\,$$^{187}\rm Os({1/ 2})^-$ has
a record-low $Q\simeq 2.47$ keV and is ``first unique
forbidden". Consequently the half-life of $^{187}\rm Re$ is long: 
4.3 10$^{10}$ years, comparable to the current age of the Universe. 
Two groups have been pursuing measurements with $^{187}\rm Re$-implanted 
calorimeters: MANU  \cite{FGatti} and
MIBETA \cite{Rhenium}. Their published limits are, respectively:
\begin{eqnarray}
m_{\nu}&<&\rm 26\, eV\, at\; 95 \% \,CL,
\nonumber\\
m_\nu&<&\rm 15.6 \, eV\, at\; 90 \%\, CL.
\label{expsRe}
\end{eqnarray}
A MIBETA Kurie plot is shown in Fig.~\ref{fig:MiBeta}.

The decay $^{163}\rm Ho(7/2)^-\!\to\! ^{163} Dy(5/2)^-$ is
an allowed ground-state to ground-state nuclear
transition.
Its half-life is a mere $\sim 4.6$ millennia.
A $Q$-value of $2.80\pm 0.08$ keV, recently obtained with a prototype
calorimeter \cite{RPK},
disagrees with the ``recommended" (and often unadvisable) 
$Q=2.555\pm 0.016$ keV \cite{Audi}. For neutrino mass measurements
it is in all cases foreseen and truly commendable to determine the $Q$-values 
independently and
very precisely with use of Penning trap techniques, which have
recently improved dramatically \cite{BNW}.

\begin{figure}
\hspace{-1.5cm}
\begin{center}
\includegraphics[width=0.45\textwidth]{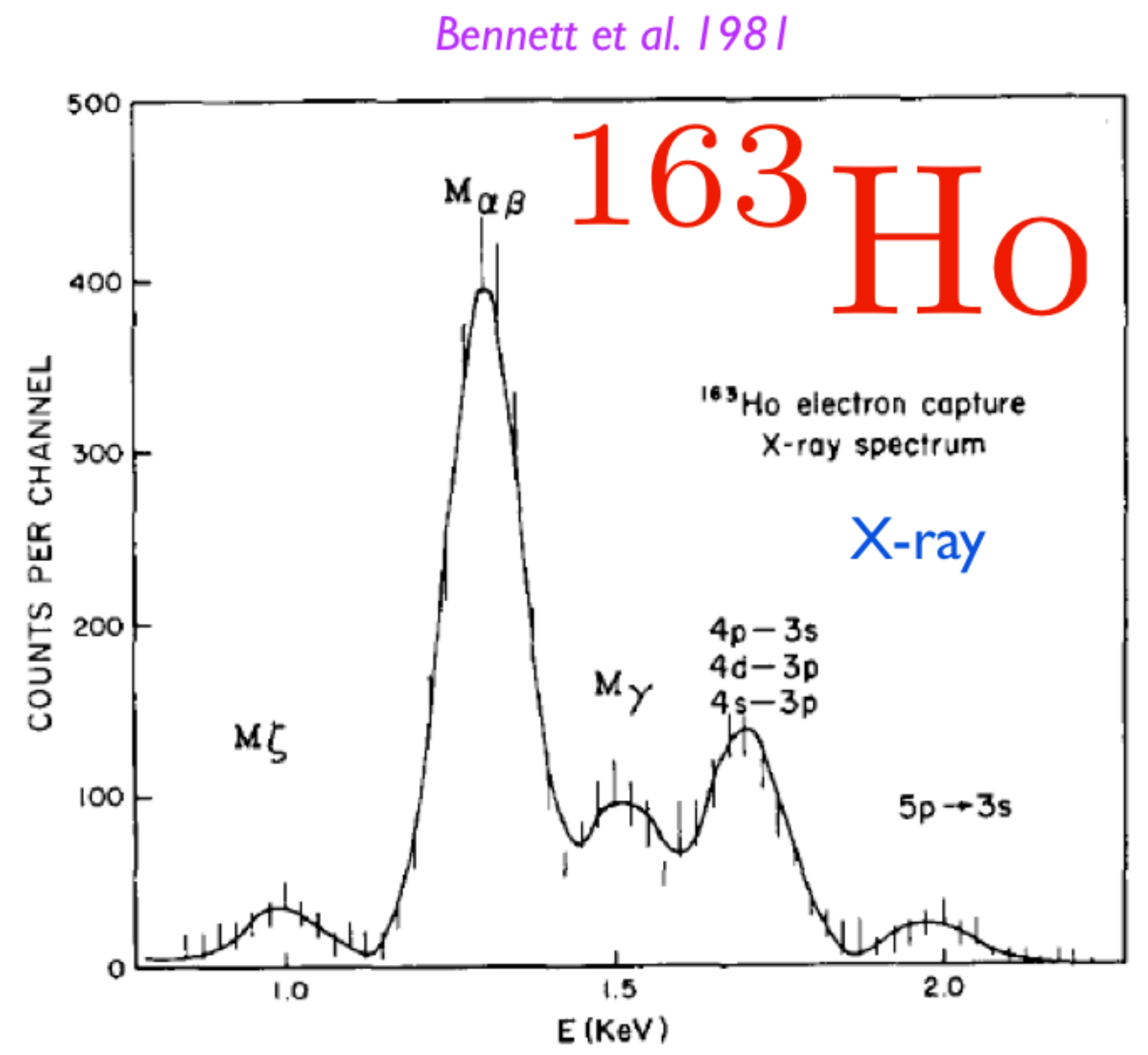}
\caption{IBEC spectrum in $^{163}\rm Ho$ decay \cite{BHN}, showing prominent
X-ray lines.
\label{fig:Bennett}}
\end{center}
\end{figure}

\begin{figure}[htbp]
\centering\includegraphics[width=0.45\textwidth]{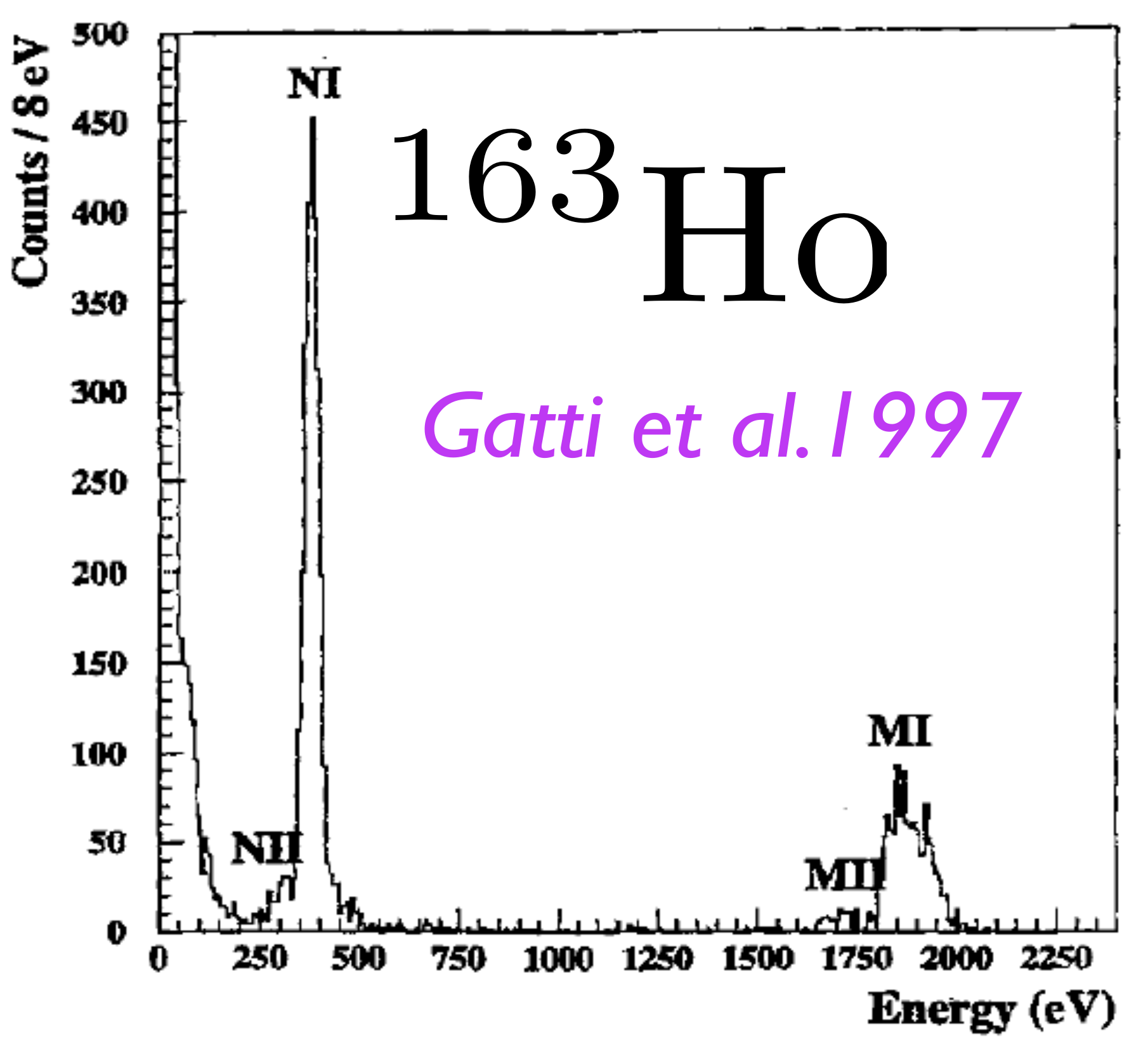}\caption{Results of an early 
$^{163}\rm Ho$ calorimetric spectrum \cite{GMSV}.
\label{fig:Gatti87}}
\end{figure}

Some early measurements with a $^{163}\rm Ho$ source \cite{BHN,SBB}
 were based on IBEC
(Internal Bremsstrahlung in Electron Capture), the first-principle theory of which
is fiendishly complex both above \cite{Glauber} and --more so-- below \cite{ADR} the energies
coinciding with X-ray resonances.  One example is shown in 
Fig.~\ref{fig:Bennett}. Other measurements were calorimetric \cite{GMSV}, 
see Fig.~\ref{fig:Gatti87}.
The most stringent of the early mass limits, from \cite{SBB} and \cite{Yasumi} were,
respectively: 
\begin{eqnarray}
m_{\nu}&<&\rm 225\, eV\, at\; 95 \% \,CL,
\nonumber\\
m_\nu&<&\rm 490 \, eV\, at\; 68 \%\, CL.
\label{OldHoExps}
\end{eqnarray}

\begin{figure}[htbp]
\includegraphics[width=0.45\textwidth]{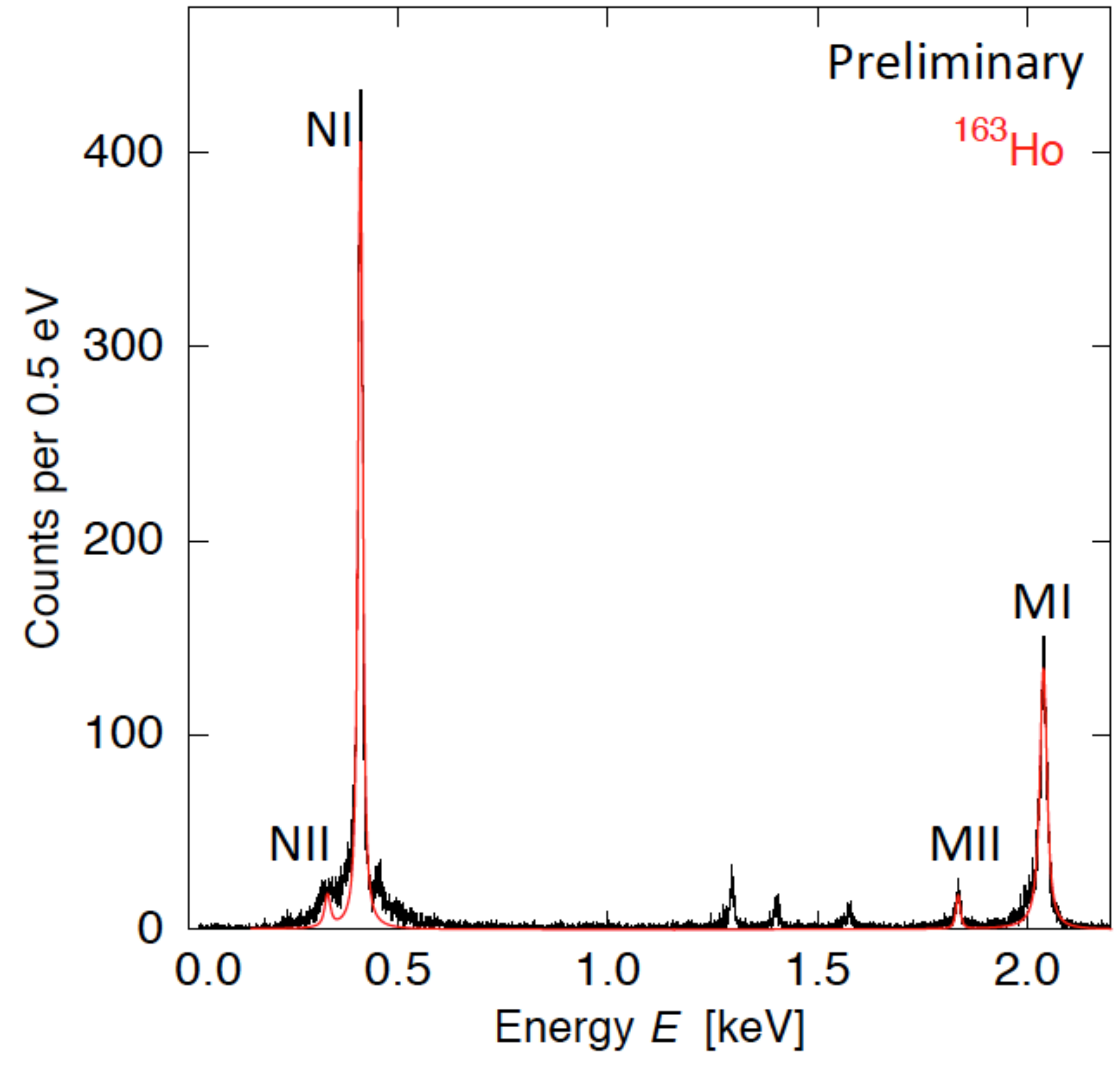}
\caption{Test results of ECHo \cite{ECHo} for the calorimetric 
spectrum of $^{163}\rm Ho$ decay. The unlabeled impurities are $^{144}\rm Pm$.
The continuous (red line) theory \cite{ADRML} is based on Eq.~(\ref{eq:resonances}).
\label{fig:Gastaldi}}
\end{figure}

The recent progress may be illustrated by comparing Fig.~\ref{fig:Gatti87} \cite{GMSV}
with the preliminary results shown in
Fig.~\ref{fig:Gastaldi}, from the incipient experiment ECHo \cite{ECHo}, which employs
MMCs (Magnetic Metallic Calorimeters). The unlabeled peaks in Fig.~\ref{fig:Gastaldi}
are due to $^{144}\rm Pm$,
an impurity accompanying  $^{163}\rm Ho$ at the implantation stage at ISOLDE-CERN,
an early test of source-preparation techniques.

\begin{figure}
\hspace{-1.5cm}
\begin{center}
\includegraphics[width=0.50\textwidth]{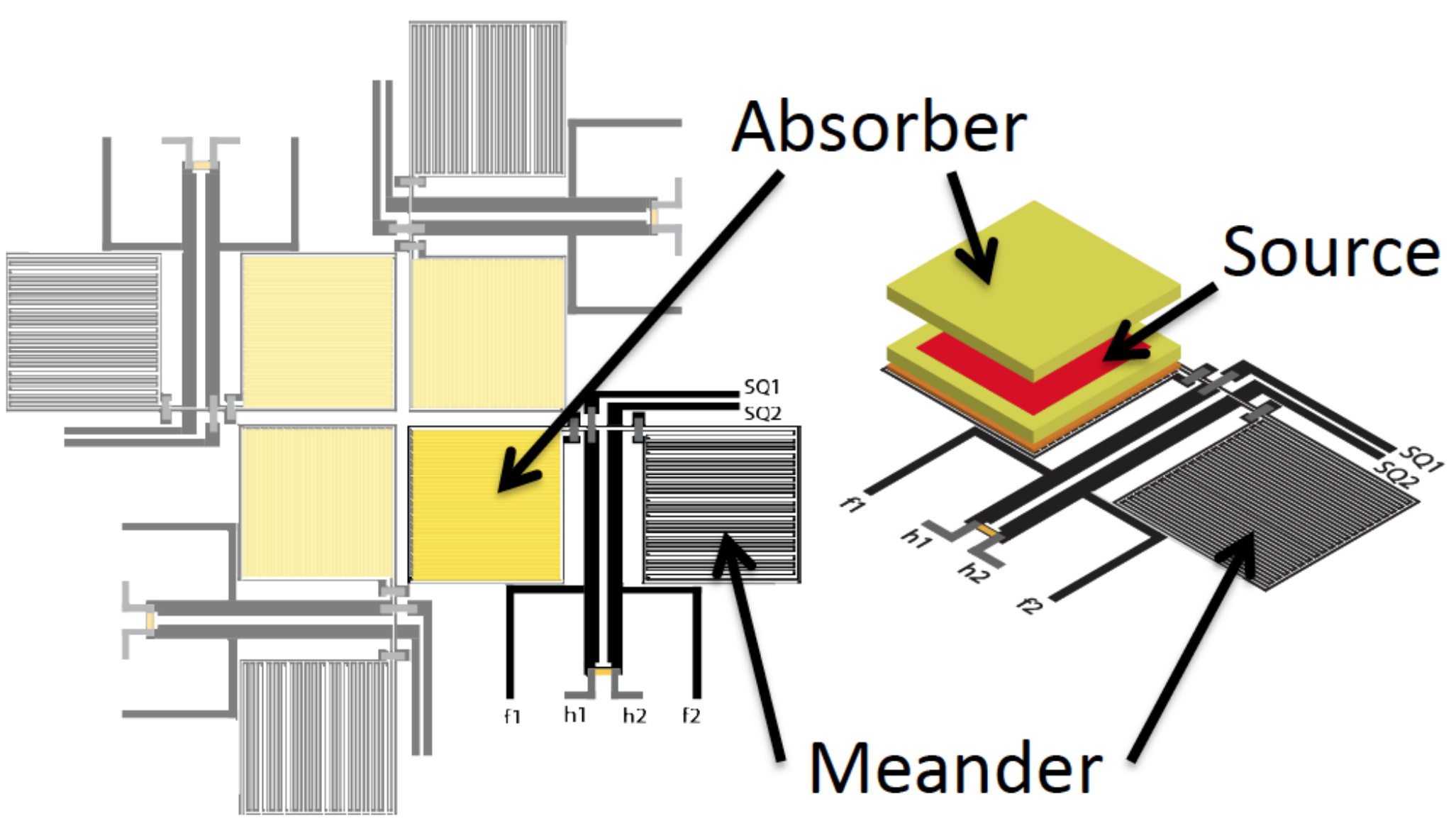}
\caption{Schematic view of an ECHo prototype. The temperature change following 
a  $^{163}\rm Ho$-decay event which deposits an energy $E_c$ in a gold absorber
is measured by the change of magnetization of a paramagnetic sensor material (Au:Er).
The ``meanders" are coupled superconducting Nb pickup coils. 
\label{fig:MMC1}}
\end{center}
\end{figure}

\begin{figure}
\hspace{-1.5cm}
\begin{center}
\includegraphics[angle=270,width=0.50\textwidth]{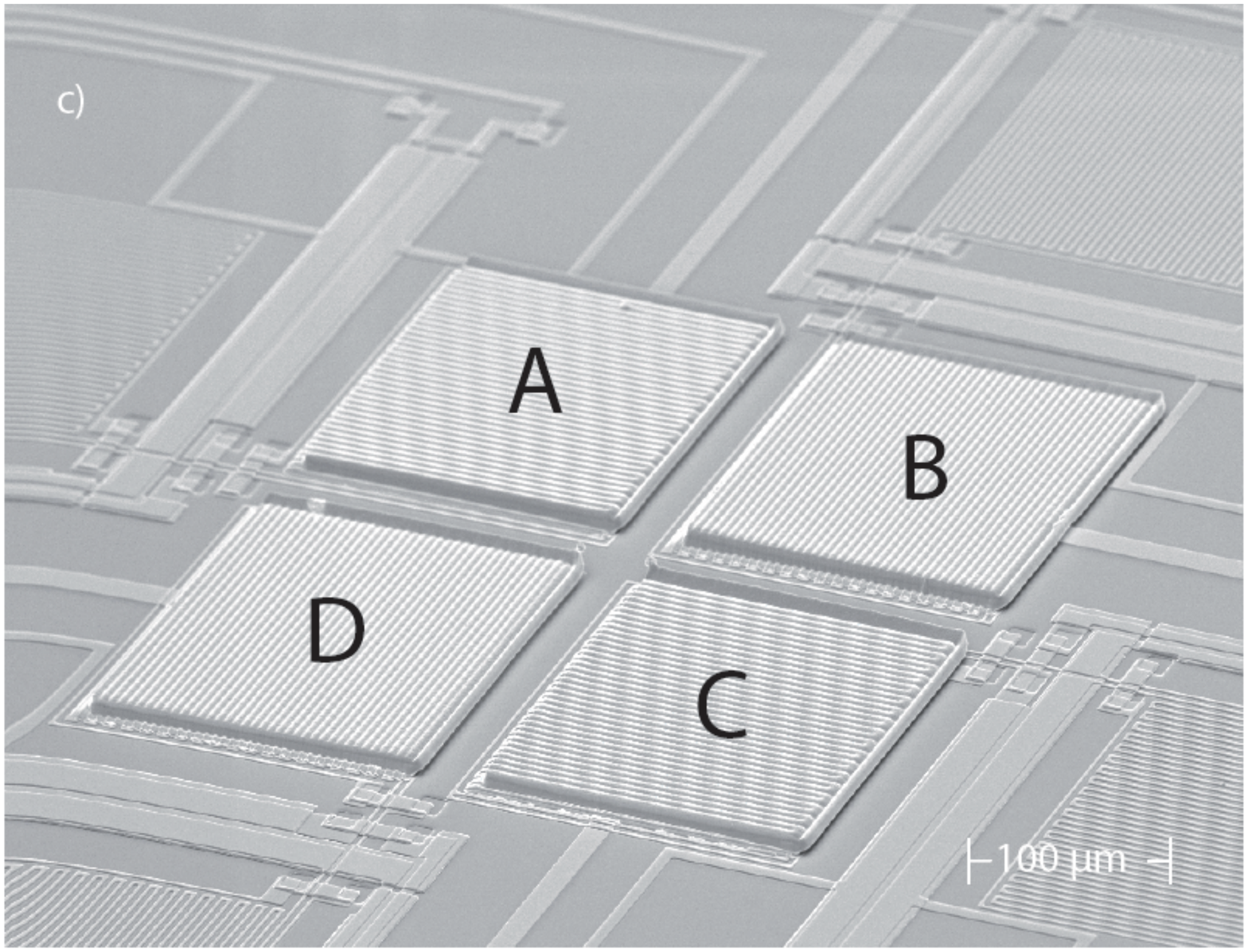}
\caption{ECHo prototype micrography. Notice the scale, implying that the
hole picture's surface is $\cal{O}$(1) mm$^2$.
\label{fig:MMC2}}
\end{center}
\end{figure}

One cannot resist the temptation of showing a scheme and a picture of the set of four 
MMCs in the $^{163}\rm Ho$ detector prototype of ECHo \cite{ECHo}:
Figs.~\ref{fig:MMC1} and \ref{fig:MMC2}.
There is satisfaction associated with the possibility of measuring a tiny
quantity --the neutrino mass-- with nano-scale detectors. Even with the associated
cryogenics and electronics, the apparatuses are still table-top.

\section{The theory of EC in $\rm{ \bold{{^{163}}Ho}}$}
\label{sec:ECHolmium}

The EC process, all by itself, does not yield any information on the neutrino mass,
or on anything else, for that matter. The mere information that ``it happened" is provided
by the fact that the daughter atom, and sometimes its nucleus, are unstable.
The hole in an atomic shell, for instance, results in observable X-rays, as the outer 
electrons cascade inwards, see Fig.~\ref{fig:Bennett}.

\begin{figure}[excitations]
\vspace{-1.5cm}
\begin{center}
\includegraphics[angle=270,width=0.50\textwidth]{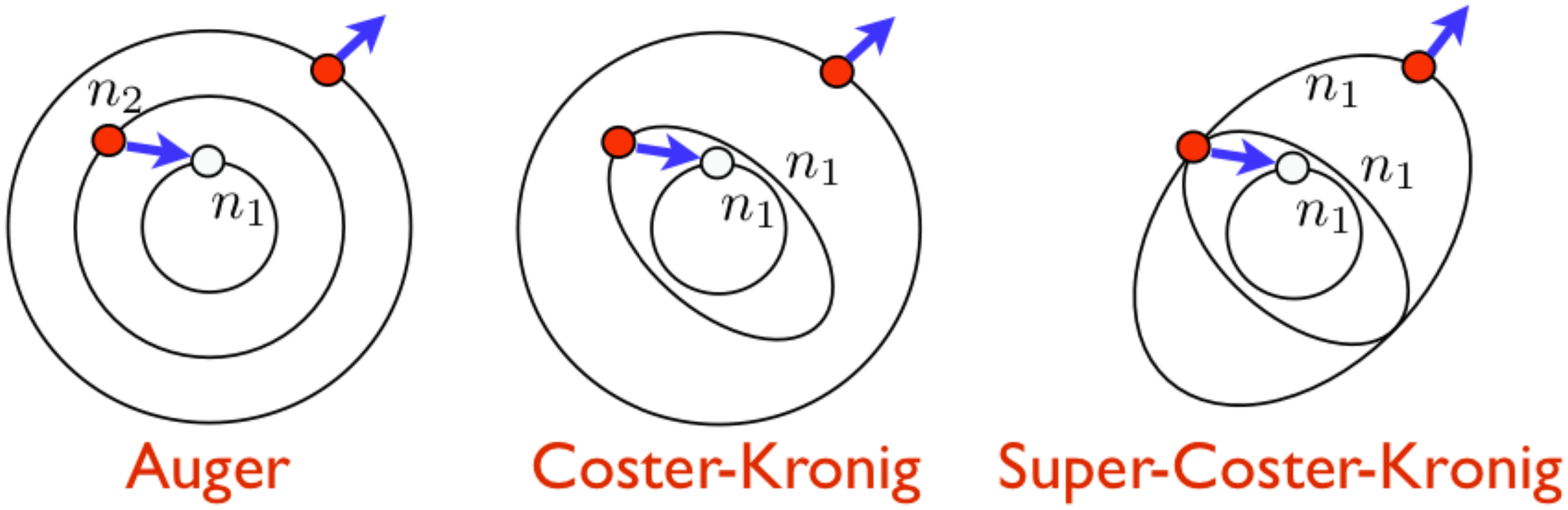}
\vspace{-1.5cm}
\caption{The names of electron-ejection de-excitations.
The holes left by EC are the hollow circles. The $n_i$ are principal
quantum numbers, which in the right-most figure all coincide.
\label{fig:Auger}}
\end{center}
\end{figure}

The measured $Q=M(^{163}{\rm Ho})-M(^{163}\rm Dy)$ is so small that EC is only 
energetically allowed from $^{163}\rm Ho$ orbitals with principal quantum number $n>2$.
The emission of X-rays from holes in such external shells is negligible compared
to that of atomic de-excitations involving electron emission (in the classical parlance, the
``fluorescence yields" are tiny). The electron-emitting transitions have more
names than interest, depending on whether or not the involved electron
orbitals have the same or different $n$. This is illustrated,
for the sake of history, in Fig.~\ref{fig:Auger}.

Let $\{n,l_j\}$ denote an atomic orbital with principal quantum number $n$, and
orbital (total) angular momentum $l(j)$. To keep its language obsolete, atomic
physics still refers to $n=1, 2, 3, 4,..$ as K, L, M, N..., to 
$\{n,l\}=\{1,0\}$, $\{2,0\}$, $\{3,0\}$, $\{4,0\}$...
as K1, L1, M1, N1..., 
to $\{n,l\}=\{1,1\}$, $\{2,1\}$, $\{3,1\}$, $\{4,1\}$... as K2, L2, M2, N2..., 
and to $l=0,1,2,...$ as S, P, D,...

For an atomic electron to be captured by its nucleus, it must
be that its wave function at the origin, $\varphi(0)$, be non-vanishing, as is the case 
for the angular momentum $l=0$, $n=3,4,5$ and $6$  
shells M1, N1, O1 and P1 in $^{163}{\rm Ho}$.
Capture from $n\rm P_{1/2}$ shells is forbidden in a non-relativistic
approximation, since $\varphi(0)=0$ for $l\neq 0$ . But the spin-orbit coupling induces
an opposite orbital parity  $n\rm S_{1/2}$ admixture of order $\alpha Z$ in the ``small"
components of the electron wave function, from which the electron can be
captured. Total angular momentum conservation forbids capture from 
$n\rm P_{j\ge 3/2}$ but for tiny corrections arising from the finite nuclear size.

All in all, in $^{163}{\rm Ho}$, energy and angular-momentum conservation allow EC
from the orbitals M1, M2, N1, N2, O1, O2 and P1, above which $\rm Ho$ runs out of electrons.

\begin{figure}[htbp]
\begin{center}
\includegraphics[width=0.45\textwidth]{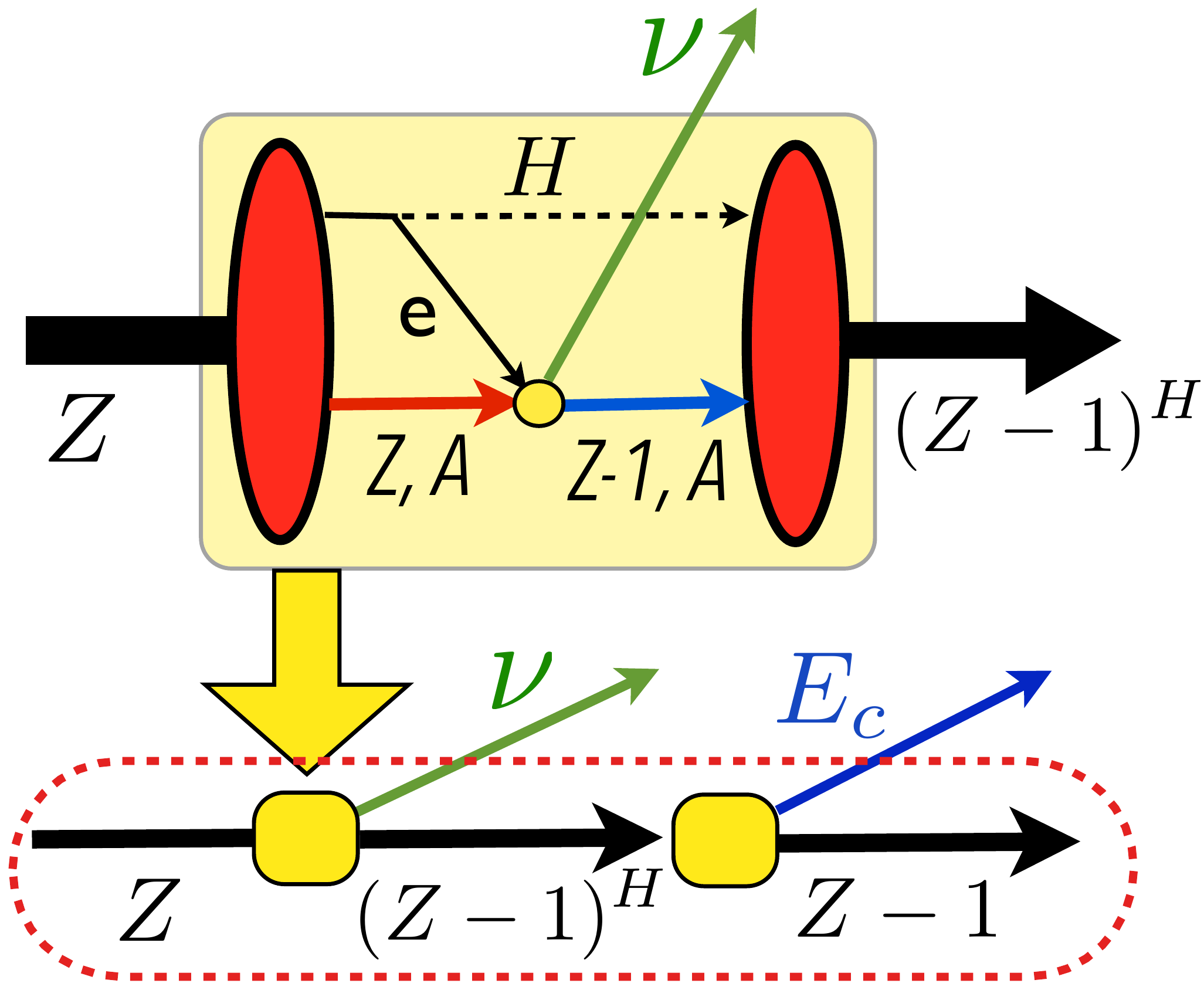}
\caption{Effective theory of electron capture \cite{ADRML}. The upper capsule embodies the
details for the decay into a daughter atom with an electron hole H. The lower
(dashed) capsule also incorporates the transition to the detector's final ground state. The
calorimetric energy $E_c$ is not meant to escape, but to be converted into a deposited-energy
signal.
\label{fig:EffectiveEC}}
\end{center}
\end{figure}

We argued in \cite{ADRML} that the matrix element for electron capture in $\rm ^{163}Ho$
may be very well approximated in an ``effective" theory extraordinarily simpler
than a first-principle QED approach \cite{ADR}. The trick consists, as in the Fig.~\ref{fig:EffectiveEC}, in 
viewing the process as a two-step one. First a two-body decay
$\rm ^{163}Ho  \rightarrow {\rm ^{163}Dy^H}+\nu_e$, with $\rm Dy^H$ any
of the relevant daughter states with a hole in the orbital H. 
Second, the
de-excitation $\rm ^{163}Dy^H  \rightarrow {\rm ^{163}Dy}+\it E_c$,
the details of which need not be specified. The double steps are to be summed over holes. 
Ignoring for the moment a series
of complications that we shall prove to be irrelevant, the
differential decay rate is then:
\begin{eqnarray}
{d\Gamma\over dE_c}&&\propto(Q-E_c)\sqrt{(Q-E_c)^2-m_\nu^2}\;
\nonumber\\
&& * \sum_{\rm H}\,
\varphi_{\rm H}^2(0)\,B_{\rm H}\,{\Gamma_{\rm H}\over 2\pi}\,{1\over (E_c-E_{\rm H})^2+\Gamma_{\rm H}^2/4}
\nonumber\\
&&{\Longrightarrow}\;\;  {\cal K}\;(Q-E_c)\sqrt{(Q-E_c)^2-m_\nu^2}\, ,
\label{eq:resonances}
\end{eqnarray}
where 
$B_{\rm H}-1$
\cite{Bambynek} is an ${\cal O}(10\%)$ correction for atomic exchange and overlap. 
The contributions to the sum in Eq.~(\ref{eq:resonances})
have a common endpoint at $E_c=Q-m_\nu$ for all H, with
${\cal K}$ (describing the spectrum near to its endpoint), a constant
to a level of precision to be discussed anon.

It is laborious to precisely compute from first principles 
the atomic parameters appearing in Eq.~(\ref{eq:resonances}).
Trusting such a calculation one may use its form, convoluted with the 
experimental resolution, to obtain information on $Q$ and even $m_{\nu}$. 
In practice it may be best to use an independently determined $Q$  \cite{BNW},
and to fit the data to the observed widths and spectral peak ratios to extract,
respectively, $\Gamma_{\rm H}$ and the ratios of the quantities 
$\varphi_{\rm H}^2(0)\,B_{\rm H}\,\Gamma_{\rm H}$. While this may
be useful in providing a fair fit to the full spectral data, it is immaterial to the
extraction information on $m_\nu$ from the endpoint spectral domain.
There, as we shall see,
the resonance-dominated atomic matrix element is practically constant.

\begin{figure}[htbp]
\hfill
\includegraphics[width=0.45\textwidth]{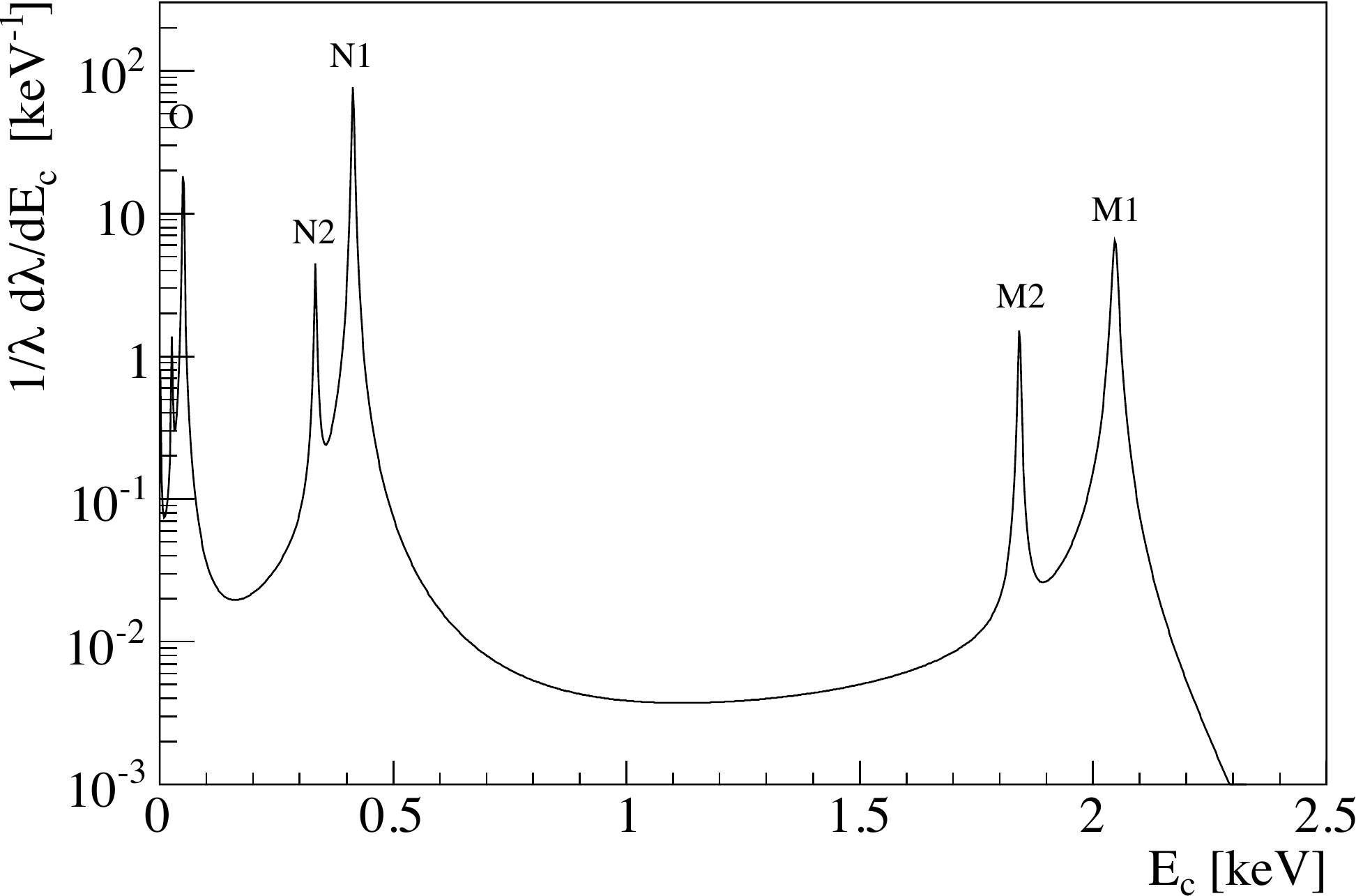}
\caption{The calorimetric spectrum of $^{163}$Ho decay \cite{ADRML,LV}.
\label{fig:LV}}
\end{figure}

A recent result based on Eq.~(\ref{eq:resonances}) is shown in
Fig.~\ref{fig:LV} \cite{LV}, for $Q=2.5$ keV,  
$B_{\rm H}=1$, $E_{\rm H}$ and $\Gamma_{\rm H}$ as
in \cite{CP} and the ratios $\varphi_{\rm H}^2(0)/\varphi_{\rm M1}^2(0)$
as tabulated in \cite{BT}.
%

\subsection{Complications?}
\label{Comp}

\subsubsection{The matrix element close to the endpoint}

\begin{figure}[htbp]
\begin{center}
\includegraphics[width=0.45\textwidth]{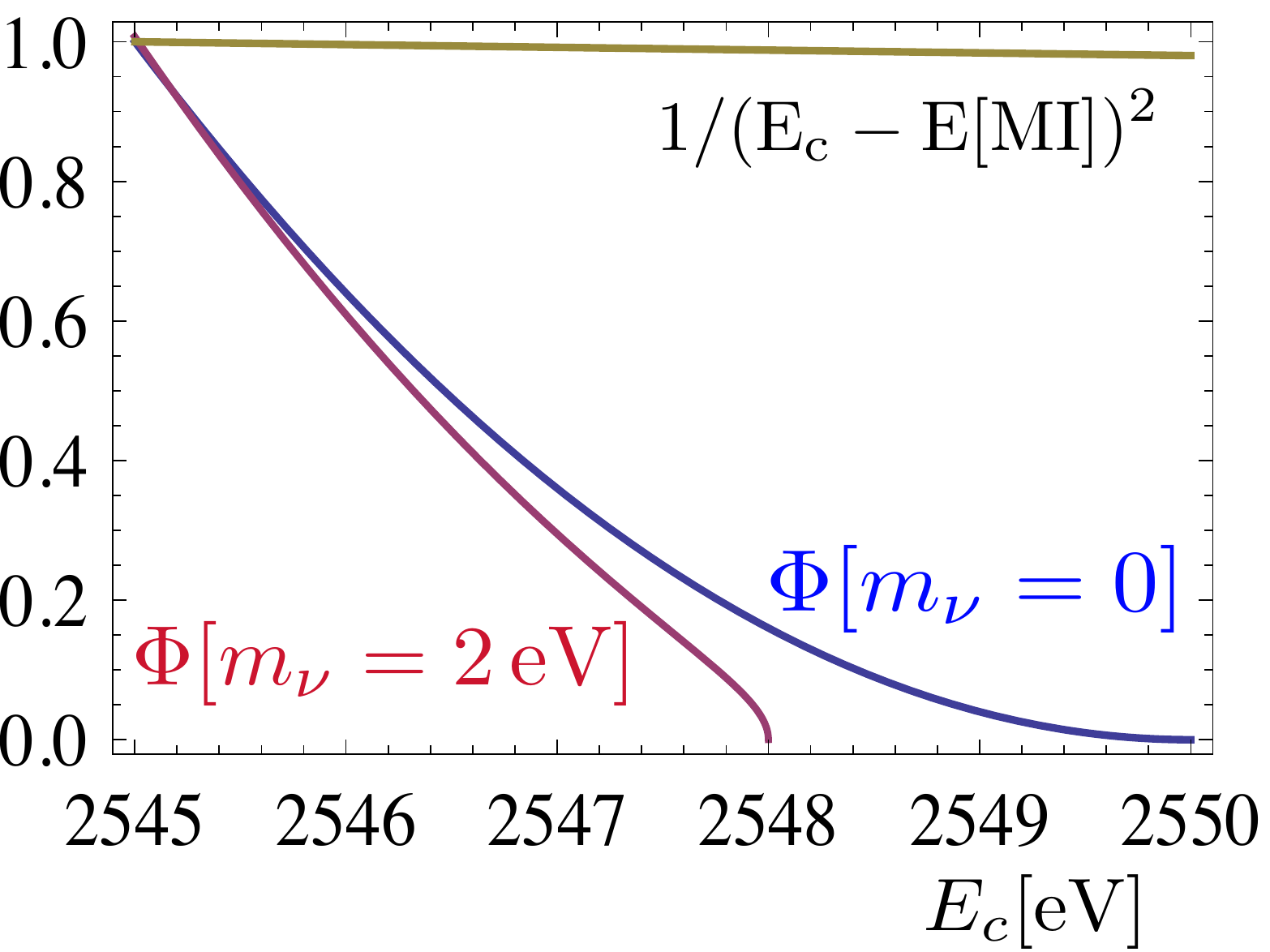}
\caption{Shapes at the endpoint of the  $d\Gamma/dE_c$ spectrum
in Ho decay, with an assumed
$Q=2.55$ keV. The $\Phi[m_\nu]$ lines are the phase-space function for two choices of
$m_\nu$. The line above them reflects the amount of energy-dependence expected
for the squared matrix element. All curves are normalized to unity at the lowest $E_c$
in the figure.
\label{fig:Endpoints}}
\end{center}
\end{figure}

An important question is the range of the largest $E_c$ values for which $\cal K$ in 
Eq.~(\ref{eq:resonances}) may in practice be taken to be a constant. The answer
depends a bit on the $^{163}$Ho decay $Q$-value, still insufficiently well measured. 
Consider the example $Q=2.55$ keV and recall that $E_{\rm H}\approx 2.05$ keV for
H = M1 Dysprosium, the state of closest energy to the endpoint. In Fig.~\ref{fig:Endpoints}
we have plotted the phase-space factor of Eq.~(\ref{eq:resonances}) 
for $m_\nu=0$ and  $m_\nu=2$ eV,
as well as the squared atomic matrix element (the sum over holes in the same equation), whose variation near the endpoint is
essentially that of the function $1/(E_c-E\rm [M1])^2$. All curves are normalized
at the lowest $E_c$ in the plot. 

The point that Fig.~\ref{fig:Endpoints} conveys is that the variation of
the matrix element, simplified or not, is governed by atomic singularities located
at the electron binding energies in Dy, as dictated by arguments as general as causality
and analyticity. The precise absolute value of the matrix element may be hard to compute,
but its variation cannot be large enough to be relevant in practice, unless the Q value
happened to fall within a few widths of a resonance. Otherwise, the figure speaks for itself.

\subsubsection{Quantum and classical effects}

The expression Eq.~(\ref{eq:resonances}) is ``classical" in two respects: 
it does not contain ``off-shell" intermediate states (the K and L, $n=1$ and 2
virtual holes) and it is a sum
of squared amplitudes and not an amplitude-sum squared. 
The K and L summands are suppressed by large energy denominators and their
negligible contribution has a squared matrix element at the endpoint
that is even flatter in energy than the one illustrated in Fig.~\ref{fig:Endpoints}.

The neglected interferences
must be small, as discussed in detail in \cite{ADRML}, because the dominant 
decay channels for Dy 
with different electronic holes are different and consequently de-cohering. The 
dominant de-excitations of $\rm H=nS,\,nP_{1/2}$, $n\!>\!2$ states are Coster-Kronig
transitions, $\rm H\to H_1\,H_2\,e$, with one of the final holes in an orbital with the same
original $n$, e.g. $\Gamma$(Coster-Kronig)$/\Gamma$(Auger) $\simeq$ 16.6, 8.6, 166, 129
for H = M1, M2, N1, N2, respectively \cite{McGuire}. In \cite{ADRML} we used the transition
matrix elements of \cite{McGuire} to estimate that the effect of interferences
is at most at the level of 1\% towards the endpoint, and totally negligible in affecting
its shape.

The neglected interferences ought to be most significant ``half-way" between two
neighboring resonances in the spectrum, far from the endpoint. A rough phenomenological
way to deal with them and with non-resonant contributions
 (complementary to doing a commendable and precise atomic calculation)
is to fit the data with two distinct widths per resonance:
one below and one above the peak. 
Once more, for the extraction of information on $m_\nu$, none of this would matter.

\subsubsection{``Instanteneous" electron ejection}

It may happen that the initial EC ``instantaneously" leaves two holes in the
daughter atom, because of the mismatch of the atomic orbitals before and
after the capture.
At first sight Eq.~(\ref{eq:resonances}) does not include this direct manner of
 production of a final state with two vacancies ($\rm H_1$ and 
$\rm H_2$) in the daughter Dy atom:
\begin{equation}
{\rm Ho}\rightarrow {\rm Dy[{H_1,H_2}]}+e^-+\nu,
\label{eq:2holes}
\end{equation}
a three-body decay with the customary extended phase space for the distribution
of electron energies. 

The previous paragraph contains a ``quantum misconception".
The ``classical" instantaneous interpretation of Eq.~(\ref{eq:2holes}) 
given in the previous paragraph is quantum-mechanically indistinguishable with
another classical interpretation of the same process.
To wit: electron capture leaving a hole in an orbital H, followed by an Auger or Koster-Kronig transition in which the hole migrates to $\rm H_2$ and an electron is ejected, 
leaving an  $\rm H_1$ hole. The two classical interpretations refer to the same initial and
final states: they are quantum-mechanically indistinct.

As an example, consider $\rm H=\rm M1$, $\rm H_2=\rm M2$, 
$\rm H_1=\rm N1$. This later process is resonant in that
the ejected electron spectrum peaks at the mass difference between Dy[M1] and Dy[M2,N1]
and the total calorimetric energy peaks at the Ho - Dy[M1] mass difference. But the process is
one of the contributions to the H = M1 peak of Eq.~(\ref{eq:resonances}).
As for all processes, its contribution  to the calorimetric energy spectrum
 extends all the way to its  endpoint at  $Q-m_\nu$, as dictated by mere energy 
conservation.

The extremely careful reader has noticed that there is one and only
one potentially relevant process not subject to the two-fold classical interpretation 
we just discussed: the instantaneous decay 
\begin{equation}
{\rm Ho}\rightarrow {\rm Dy^{M1,N1}}+e^- +\nu.
\label{eq:2darnedholes}
\end{equation}
This decay is possible thanks to the slightly incomplete overlap between the
wave function of the N1 electron in Ho and in Dy with an M1 vacancy. The
charge that the N1 electrons feel in these two atoms
is not  the same, since electron screening is not perfect: as ``seen" by an N1 
electron, the M1 electron absent in the daughter disprosium does not
completely compensate for the absence of a proton in the Dy nucleus,
relative to Ho.

The decay channel of Eq.~(\ref{eq:2darnedholes}) is non-resonant and has
a negligible rate relative to the resonant processes described by Eq.~(\ref{eq:resonances}).
Moreover, the shape of its calorimetric endpoint is, once again, that of the last
line of Eq.~(\ref{eq:resonances}). It is instructive to prove these two points in some detail.


To get a very rough order of magnitude of the probabilities $P_s$ 
for an N1 electron staying in place --and $P_e$ for one of the two
N1 electrons being instantaneously ejected-- in the capture of 
an M1 electron, one may use Coulomb wave functions to get:
\begin{eqnarray}
&&P_s=|\langle \Psi_{\rm N1}(Z) | \Psi_{\rm N1}(Z-1)\rangle|^2=
1-\frac{33}{4
   Z^2}+{\cal O}\left[\frac{1}{Z}\right]^3,
   \nonumber\\
&&  P_e \approx 2\,(1-P_s)\approx 
\frac{33}{2 Z^2}\approx 3.6\times 10^{-3},\,{\rm for}\;Z=68.
\label{Approx}
   \end{eqnarray}
%
   

A better estimate of $P_e$ would be obtained with use of explicitly computed
Ho and Dy atomic wave functions, but that would be an overestimate
since the charge seen by the Dy N1 electrons  is not
$Z-1$. Since there are no computed, readily available wave
functions for Dy with an M1 hole, we shall evaluate all effects of screening.

A Coulomb wave function for M1 or N1 electrons with $Z=68$ is also not a good
choice since, again, it does not reflect the charge screening from the 
rest of the electrons, particularly the $n=1,2$ inner ones. To correct for
this as we did in similar calculations in \cite{EEEC},
let us introduce a Coulombic $Z_{\rm eff}(n)$ giving the same
mean orbital radii as in an elaborate calculation with Roothaan-Hartree-Fock atomic wave functions \cite{McLean}.
The results are:
\begin{eqnarray}
&&Z_{\rm eff}^{\rm Ho}(\rm M1)\approx 54.91, \;\; Z_{eff}^{\rm Ho}(\rm N1)\approx 43.22,
\nonumber\\
&&Z_{\rm eff}^{\rm Dy}(\rm M1)\approx 53.95, \;\; Z_{eff}^{\rm Dy}(\rm N1)\approx 42.42.
\end{eqnarray}
From the  Coulombic charge distributions, we can make an estimate of the charge,
$\alpha$, by which an
M1 hole in Ho EC would screen
an N1 electron in the daughter Dy:
\begin{eqnarray}
\alpha\!\!&\!=\!&\!\!
\!\int_0^\infty \!\!\!dr^3\int_r^\infty\!\!\! dr'^3
|\Psi_{\rm M1}(Z_{\rm eff}^{\rm Ho}(\rm M1),r)|^2   
|\Psi_{\rm N1}(Z_{\rm eff}^{\rm Dy}(\rm N1,r')|^2 
\nonumber\\
&\approx& 0.9649,
\end{eqnarray}
indicating a very large screening, i.e.~$\epsilon\equiv 1-\alpha\ll 1$.
The use of Hartree-Fock wave functions, as opposed to $Z_{\rm eff}$
approximations, gives results for quantities involving wave function
overlaps, such as $\epsilon$, that differ by no more than a factor
of ${\cal O}(1)$ \cite{EEEC}. The advantage of the Coulombic approximation
is that it makes the underlying physics very transparent, as in the next paragraph.

Define $\tilde Z\equiv Z_{\rm eff}^{\rm Ho}(\rm N1)$ to write the
probabilities introduced prior to Eq.~(\ref{Approx}), now fully corrected for screening:
\begin{eqnarray}
      P_s&=&|\Psi_{\rm N1}(\tilde Z|
       \Psi_{\rm N1}(\tilde Z-\epsilon)\rangle|^2
      \nonumber\\
      &=&
1-\frac{33 \,\epsilon ^2}{4
   {\tilde Z}^2}\left\{1+{\cal O}\left[\epsilon,{1/\tilde Z}\right]\right\} ,
   \nonumber\\
    P_e &\simeq&2\,(1-P_s)\approx \frac{33 \,\epsilon ^2}{2 {\tilde Z}^2}=1.08\times 10^{-5}.
   \end{eqnarray}
There is a slight enhancement relative to the naive estimate of $P_e$ in Eq.~(\ref{Approx}), 
due to the change $Z\to \tilde Z$, and an enormous reduction due to the factor
$\epsilon^2\simeq 1.2 \times 10^{-3}$.

The conclusion is that the instantaneous decay of Eq.~(\ref{eq:2darnedholes})
is totally negligible ($P_e\!\ll\! 1$), relative to the M1 capture contribution to Eq.~(\ref{eq:resonances}).
Moreover, the shape of the corresponding squared matrix element close to the endpoint
is the same as in that equation, since the extra factor $p_e$ of electron-ejection phase
space is compensated by the Fermi function $F(E_e)$, mentioned in $\S$ \ref{reminder},
which at the relevant
very low electron energies behaves as $1/p_e$. The atomic matrix
element for the negligible process of Eq.~(\ref{eq:2darnedholes}) contains a non-resonant
denominator $E_e+E(\rm N1)$ that would have an even lesser shaping effect than the one
illustrated in Fig.~\ref{fig:Endpoints}. Naturally, all this discussion ought to be
complemented by an analogous one with ${\rm M1}\!\leftrightarrow\!{\rm N1}$, 
the results of which are equally negligible.

\subsubsection{Two experimental issues}

Implicit in the calorimetric considerations of $\S$ \ref{sec:CP}
 is the hypothesis that the de-excitation time
of $\rm ^{163}Dy^H$ to its ground state is faster than the $\sim 0.1\,\mu$s
rise-time for signals already achieved in prototype MMCs \cite{Gastaldo}
(the signal decay-time is much longer). Atomic excited states having inverse
widths of ${\cal O}(1)$ eV$^{-1}\sim 10^{-15}$ s, this seems to be a safe
expectation, barring the existence of unforeseen metastable final states.

A more serious consideration is the possibility that Ho atoms in the detector
be bound, not to one type of chemical neighbourhood, but to more than one.
That would mean that the calorimeter is a sum of detectors with
$Q$-values that may differ by an eventually significant amount \cite{Rabin}.

\section{Double electron capture}

\begin{figure}[htbp]
\includegraphics[width=0.45\textwidth]{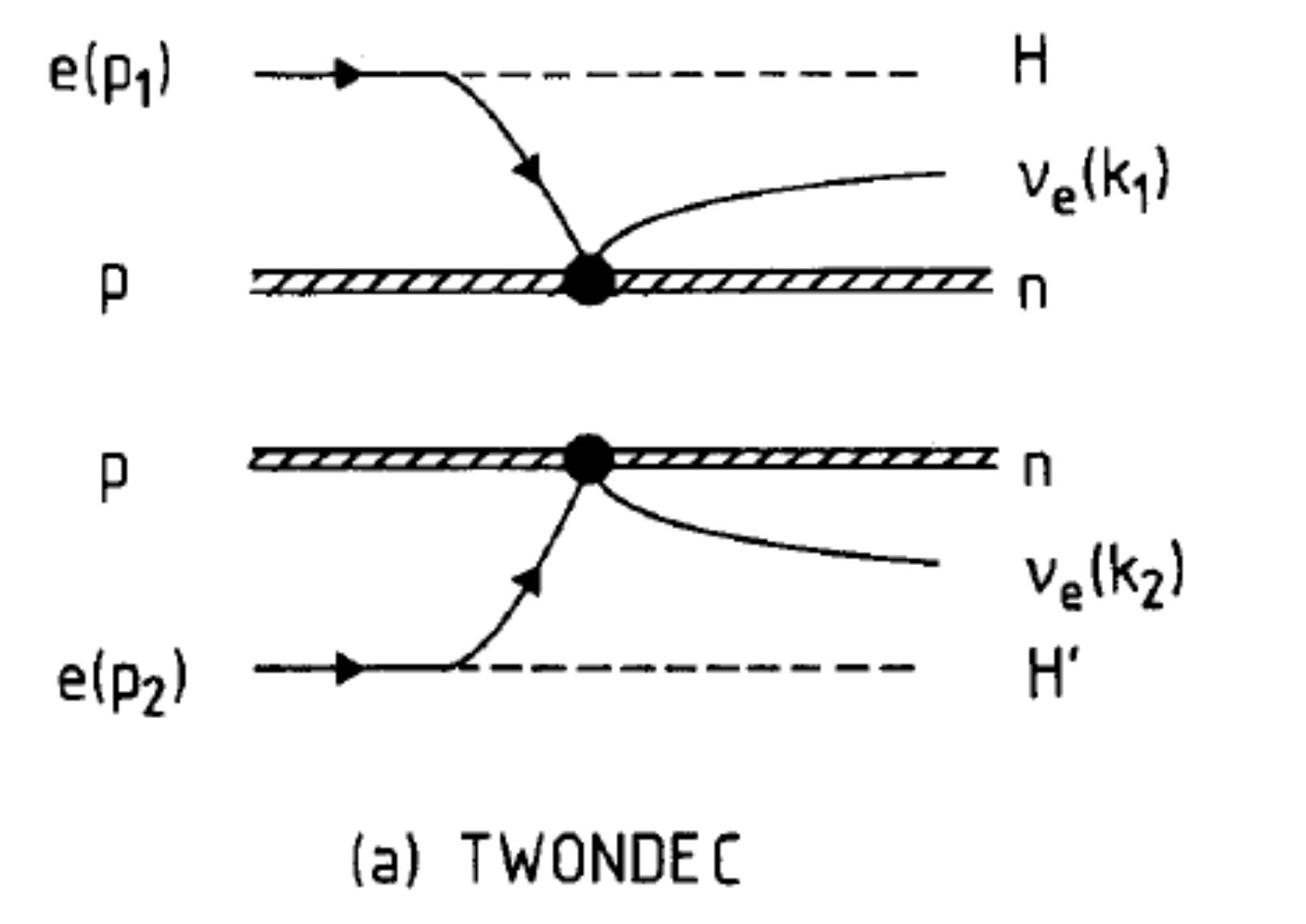}\\
\includegraphics[width=0.45\textwidth]{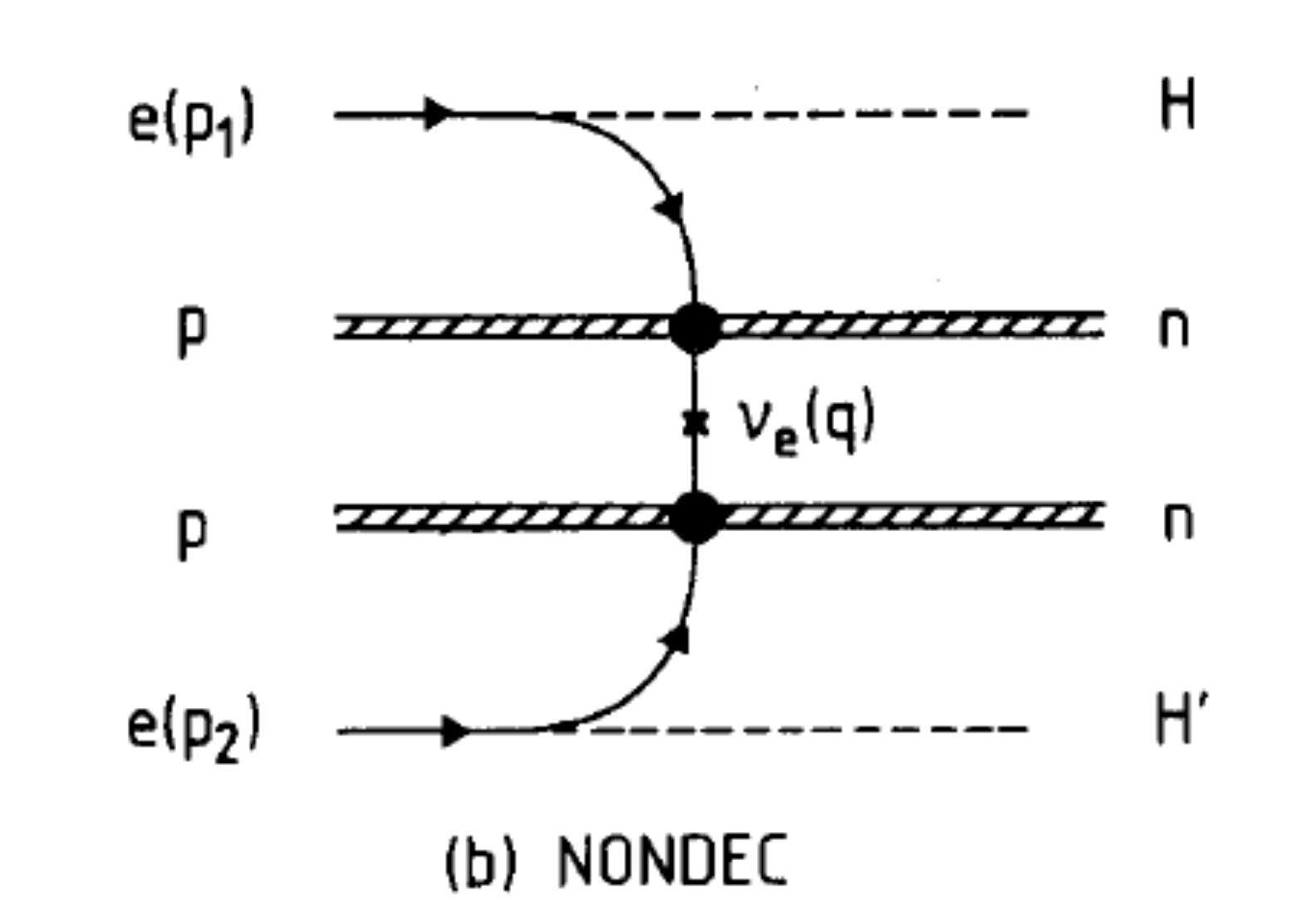}
\caption{Feynman diagrams for: (a) Two-neutrino double EC (TWONDEC). (b)
No neutrino double EC (NONDEC) \cite{BDJ}.
\label{fig:DEC}}
\end{figure}

Neutrino-less double $\beta$-decay has
an EC analog: neutrino-less double electron capture \cite{BDJ}. 
The Feynman diagrams for double EC (DEC) processes, 
with two or no outgoing neutrinos, are shown in Fig.~\ref{fig:DEC}.

\begin{figure}[htbp]
\includegraphics[width=0.5\textwidth]{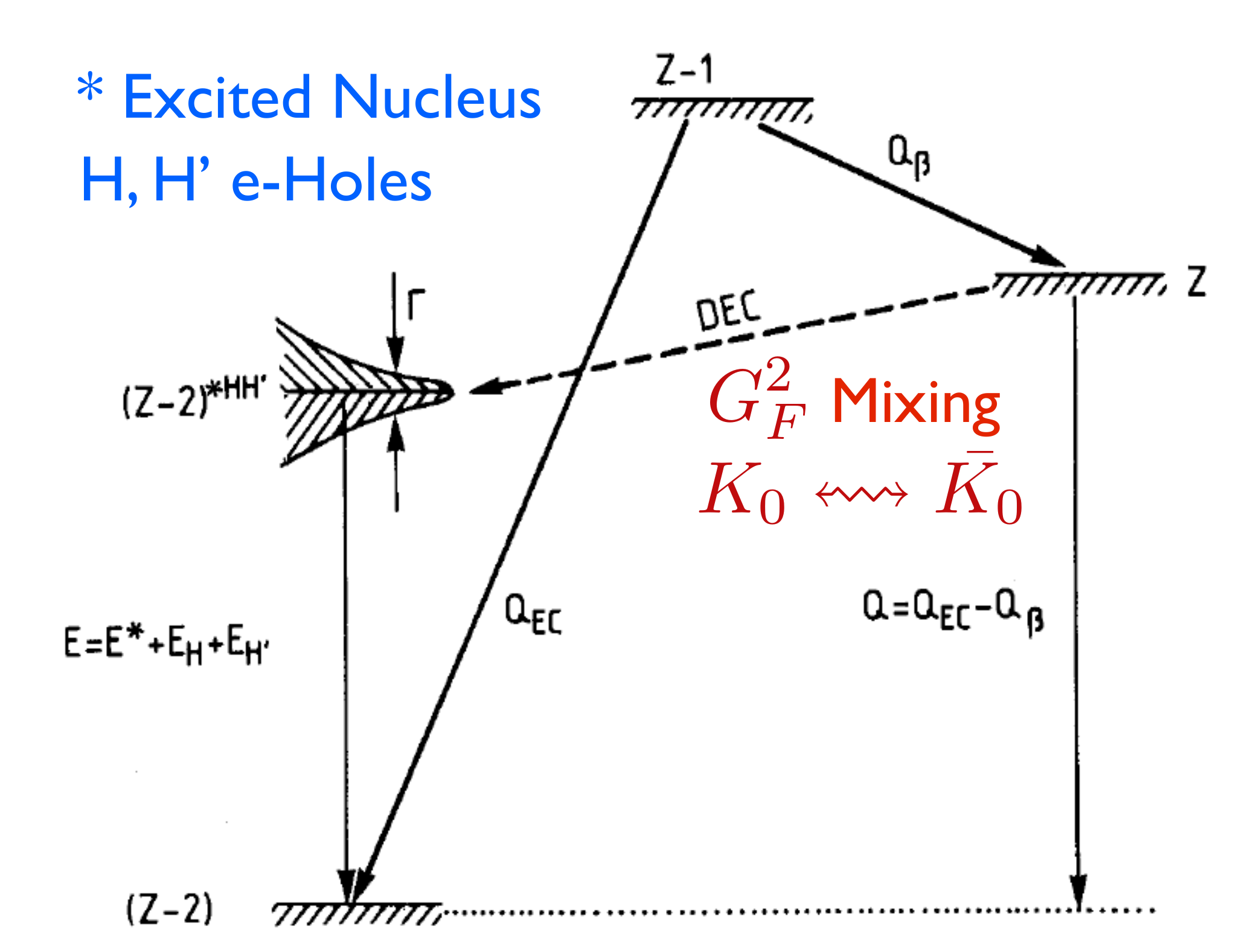}
\caption{Level structure of allowed DEC.
The symbols and the analogy with $K^0\leftrightarrow \bar{K}^0$
mixing are explained in the text.
\label{fig:DEClevels}}
\end{figure}

The level structure and energetics of the atoms that undergo
DEC are illustrated in Fig.~(\ref{fig:DEClevels}). An ``intermediate"
atom of nuclear charge $Z-1$  $\beta$-decays to $Z$ with a $Q$-value
$Q_\beta$, and EC-decays to $Z-2$ with a $Q$-value $Q_{\rm EC}$.
The mass difference between the ground-states of $Z$ and $Z-2$ is 
$Q=Q_{\rm EC}-Q_\beta$. DEC from $Z$ to $Z-2$ is energetically allowed if $Q$ is greater
than the sum $E_{\rm H}+E_{\rm H'}$ of binding energies of the
electron orbitals vacated by the double capture. The figure includes the
possibility that the transition $Z\to Z-2$ be to an excited daughter
nucleus of energy $E^*$ above the ground state. In that case, requiring
$Q>E_{\rm H}+E_{\rm H'}+E^*$, the
decay $(Z-2)^*\to Z-2+\gamma$ may constitute
an additional signature:  the nuclear $\gamma$-ray.

Double $\beta$-decay experiments measure the spectrum of the
sum of energies, $E_T$, of the two outgoing electrons. The dominant
process has two outgoing neutrinos. The searched-for neutrino-less
process would appear as a narrow peak, of width compatible with the resolution,
at the endpoint of the $E_T$ spectrum. The two-neutrino decay
constitutes a significant irreducible background. 

A potential advantage 
of neutrino-less DEC is that its corresponding two-neutrino irreducible background
is, relative to the signal, rendered negligible by the three-body phase-space
suppression of the background. A calorimetric measurement would only see
the peak at $E=E_{\rm H}+E_{\rm H'}$, if the daughter nucleus is stable. Otherwise,
with a nuclear $\gamma$-ray also involved, one can think of many other possibilities,
at least if ``one" is a theorist.

\subsection{The theory on neutrino-less DEC}

\begin{figure}[htbp]
\includegraphics[width=0.5\textwidth]{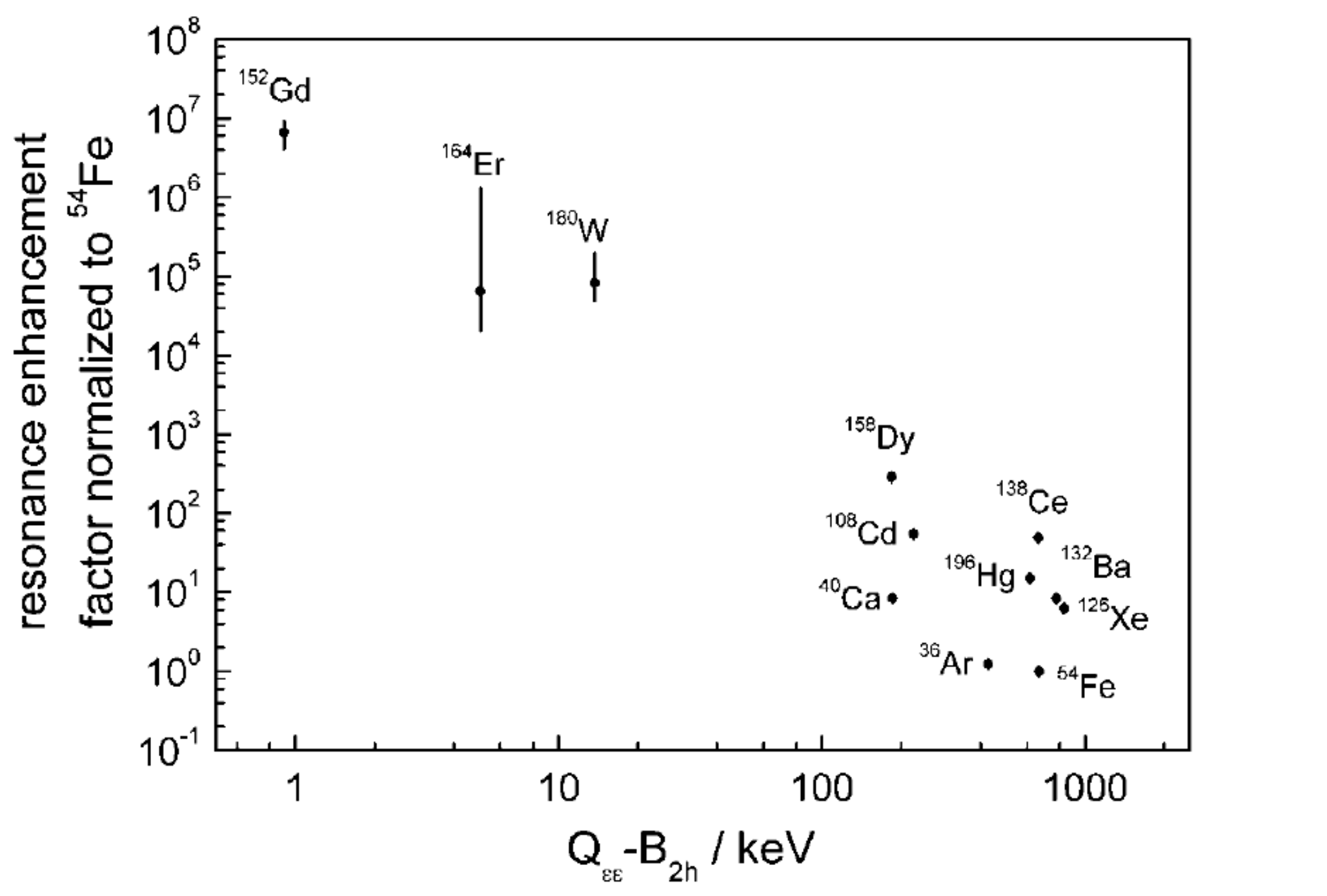}
\caption{Enhancement factors of no-neutrino DEC isotopes, relative to $^{54}\rm Fe$
\cite{BNW}. Notice the result for $^{152}\rm Gd$.
\label{fig:DECEnhanced}}
\end{figure}

In \cite{BDJ} we studied in detail neutrino-less DEC and the cases
for which, in analogy with EC in $^{163}\rm Ho$, the process could be very significantly
resonant-enhanced. At the time, the available information on the relevant
$Q$-values was not very precise, and of the dozen nuclides we
selected, $^{152}\rm Gd \to ^{152}\rm Sa$, a 
$0^+\to 0^+$(ground-state) nuclear transition, now appears to be 
an optimal candidate \cite{BNW}. To wet the reader's appetite for a reminder of the underlying
theory, I show in Fig.~\ref{fig:DECEnhanced} the resonant-enhancement factors
for various DEC isotopes \cite{BNW}, normalized to the quintessential DEC-decaying one,
$^{54}\rm Fe$. Most known relevant DEC cases are double-K orbital captures, an
exception being the double-L capture in $^{164}\rm Er\to{^{164}\rm Dy}$, also shown
in the figure.

The neutrino mass parameter $\bar m_\nu$ relevant to neutrino-less DEC or 
double $\beta$-decay are obviously the same:
\begin{equation}
\bar m_\nu=\sum_i (U^*_{ei})^2\,m_i.
\end{equation}

The theory of no-neutrino DEC can be easily understood by analogy with
$K^0\!\leftrightarrow\! \bar K^0$ mixing \cite{BDJ}. The parent atom virtually mixes, 
also with an amplitude of ${\cal O}(G_F^2)$, with the daughter one. 
Having two electron holes, the unstable daughter ``leaks".
Consider the most general case including the possibility that the
daughter nucleus be unstable and let $E\equiv E_{\rm H}+E_{\rm H'}+E^*$ and
$\Gamma\equiv  \Gamma_{\rm H}+\Gamma_{\rm H'}+\Gamma^*$.
Let $\Delta M$ be the non-diagonal element of the mass matrix of
the parent-daughter system.
This suffices to write, for the lifetime $\tau$ and half-life $T_{1/2}$ of no-neutrino DEC:
\begin{equation}
{1\over \tau}\equiv {\ln 2\over T_{1/2}}={\Gamma\over (Q-E)^2+\Gamma^2/4}\, (\Delta M)^2.
\label{resonance}
\end{equation}
This expression transparently describes the name of the game: to
find cases with $Q-E$ as small as possible and $\Delta M$ not
suppressed by angular momentum conservation (i.e.~not involving
a forbidden nuclear transition).

The form of $\Delta M$ is simple, at least for the 
$^{112}{\rm Sc}\!\to\! {^{112}\rm Cd}^*$, $0^+\!\to\! 0^{+}$ transition to an excited
nucleus with $E^*= 1.871$ MeV we favoured in \cite{BDJ}, or the
decay $^{152}{\rm Gd}\!\to\! {^{152}\rm Sm}$, a $0^+\!\to\! 0^{+}$ (ground state)
case favoured after the precise $Q$-value determinations of \cite{BNW}. To wit:
\begin{equation}
\Delta M = {1\over 4\pi R_N}(G_F\cos\theta_C)^2
\varphi_{\rm H}(0)\varphi_{\rm H'}(0)\,
g_A^2\,  |{\cal M}|\,\bar m_\nu,
\end{equation}
where $R_N=\langle R^2 \rangle^{1/2}$ is the root-mean-squared nuclear radius
and $|{\cal M}|$ (a number) is the rest of the nuclear matrix element.
The precise calculation of  $|{\cal M}|$  is not simple, but
ample progress has been recently made \cite{nuclear}.

An up-to-date calculation of the neutrino-less half-life of $^{152}{\rm Gd}$ 
\cite{Eliseev,Fang} results in:
\begin{equation}
T_{1/2}= 10^{26}\, |1\; {\rm eV}/ \bar m|^2\;\rm  years,
\end{equation}
in the ballpark of the limits in searches for neutrino-less  double 
$\beta$-decay.  This opens up the possibility of using this isotope
in a calorimetric experiment \cite{Eliseev}.

\section{The sum of neutrino masses}

\begin{figure}[htbp]
\includegraphics[width=0.45\textwidth]{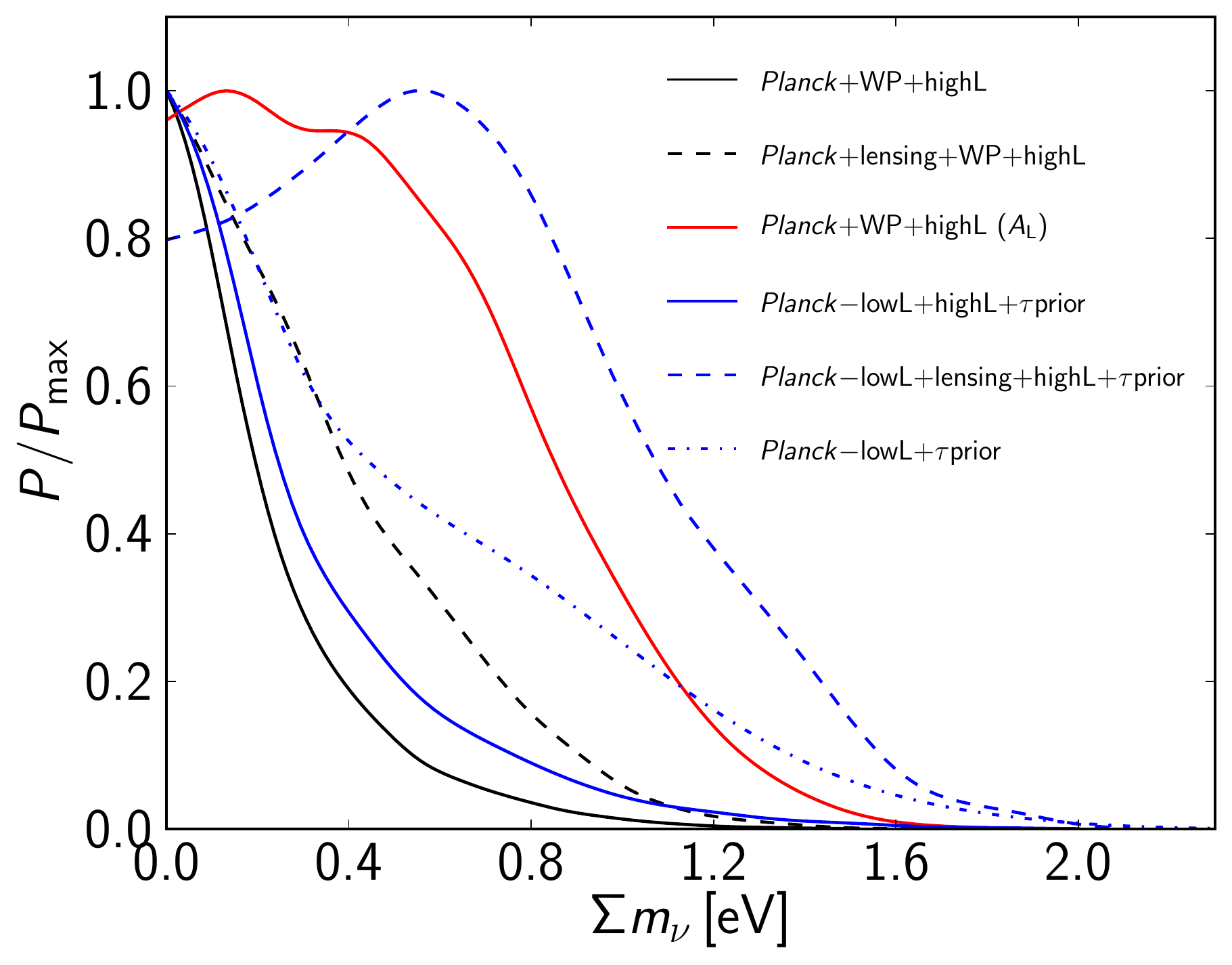}
\caption{The normalized $\sum m_\nu$ probability distributions and various combinations of
data and priors. For details, see \cite{Planck}.
\label{fig:Planlmnu}}
\end{figure}

Cosmological observations place upper limits on $\sum m_{\nu }$, the sum of masses of
three light neutrinos. These limits are quite model dependent and
vary strongly with the data combination adopted, as illustrated in Fig.~\ref{fig:Planlmnu},
reproducing the recent Planck satellite results \cite{Planck}
for the normalized $\sum m_\nu$ probability distributions resulting 
from various combinations of input data and priors.

From their data, the Planck collaboration quotes 95\% CL limits on $\sum m_\nu$
 that range from 
0.66 eV to 1.31 eV, depending on the chosen data and priors \cite{Planck}.

The neutrino oscillation data are not subject to the complications and subtleties
of cosmological analyses. They
provide {\it lower} limits on the sum of neutrino masses
for three neutrino species:
\begin{eqnarray}
\sum_i m_{\nu i}&>&0.06\;\rm eV,
\nonumber\\
\sum m_{\nu i}&>&0.1\;\rm eV,
\end {eqnarray}
for a normal-hierarchy and for an inverted-hierarchy of masses, respectively \cite{GG}.
These results do establish a target for direct-measurement experiments to aim at.

\section{Conclusions}

I have argued that the theory required to analyze the results of past and future 
calorimetric measurements of the electron neutrino mass is simple and precise enough.
This statement applies to
single electron capture processes, as well as the neutrino-less 
double electron capture ones that are
relevant only if neutrinos are Majorana.

The calorimetric measurements in the case of allowed nuclear transitions, that I have
extensively discussed for EC in $^{163}\rm Ho$, should not have any of the atomic
and molecular complications present in spectrometric $\beta$-decay experiments.
The sophisticated calorimeters, thermometers, read-out electronics
 and source-preparation tools required
for these experiments to be quantitavely competitive are being developed only recently, with decades of delay
relative to $\beta$ decay. But micro-calorimeters have the irresistible aesthetic advantage
of being tiny contraptions to measure a tiny mass.

Enormously improved data on the $Q$-values relevant to neutrino-less double
electron capture  point to $^{152}\rm Gd$ as the optimal source, for
which calorimetric measurements ought to be possible and theoretically very clean.

The masses of neutrinos, their weak mixing angles, the flux of cosmic rays, the density
and height of the atmosphere, the lifetimes of pions and muons, 
the inner temperature of the Sun,
its radius and density profile, the size and density of the Earth, the parameters required
for the Sun and nuclear reactors to function... have all been chosen by nature --just so-- that we can measure the
properties of neutrinos in Earth-based experiments. 

Thus, there is ample reason to hope that the previous paragraph can be made extensive
to laboratory measurements of the mass of the electron-neutrino and of the
Dirac or Majorana character of these particles.

\vspace{.3 cm}

{\bf Aknowledgements} 

I am very indebted to Maurizio Lusignoli for our collaboration,
his suggestions and his very careful reading of the manuscript.

%
%
%
\end{document}